\documentclass[english,final, smallcondensed, epjp3]{svjour3}
\usepackage[T1]{fontenc}
\usepackage[latin9]{inputenc}
\usepackage{geometry}
\usepackage{float}
\usepackage{url}
\usepackage{amsmath}
\usepackage{graphicx}
\usepackage{esint}

\makeatletter

\newcommand{\noun}[1]{\textsc{#1}}

\journalname{European Physical Journal Plus}
\usepackage{babel}
\usepackage{tikz-qtree}
\usetikzlibrary{positioning,shadows,arrows,trees,shapes,fit}
\usepackage{bm}
\usepackage[final]{microtype}
\usepackage{cite}
\usepackage{xcolor}
\usepackage[unicode=true,bookmarks=true,bookmarksnumbered=false,
bookmarksopen=false, breaklinks=true,pdfborder={0 0 0},
pdfborderstyle={},backref=false,colorlinks=true,citecolor=blue,linkcolor=blue]{hyperref}

\makeatother

\usepackage{babel}
\begin{document}

\title{Covariant theory of gravitation in the framework of special relativity\thanks{Published in Eur. Phys. J. Plus (2018) 133: 165. http://doi.org/10.1140/epjp/i2018-11988-9.}}

\author{R. S. Vieira \and H. B. Brentan}

\institute{Universidade Federal de São Carlos. Caixa postal 676, CEP. 13565-905,
São Carlos-SP, Brasil. \\ \email{rsvieira@df.ufscar.br \and brentan@df.ufscar.br}}

\maketitle
\abstract{In this work, we study the magnetic effects of gravity
in the framework of special relativity. Imposing covariance of the
gravitational force with respect to the Lorentz transformations, we
show from a thought experiment that a magnetic-like force must be
present whenever two or more bodies are in motion. The exact expression
for this gravitomagnetic force is then derived purely from special
relativity and the consequences of such a covariant theory are developed.
For instance, we show that the gravitomagnetic fields satisfy a system
of differential equations similar to the Maxwell equations of electrodynamics.
This implies that the gravitational waves spread out with the speed
of light in a flat spacetime, which is in agreement with the recent
results concerning the gravitational waves detection. We also propose
that the vector potential can be associated with the interaction momentum
in the same way as the scalar potential is usually associated with
the interaction energy. Other topics are also discussed, for example,
the transformation laws for the fields, the energy and momentum stored
in the gravitomagnetic fields, the invariance of the gravitational
mass and so on. \\
 We remark that is not our intention here to propose an alternative
theory of gravitation but, rather, only a first approximation for
the gravitational phenomena, so that it can be applied whenever the
gravitational force can be regarded as an ordinary effective force
field and special relativity can be used with safety. To make this
point clear we present briefly a comparison between our approach and
that based on the (linearized) Einstein's theory. Finally, we remark
that although we have assumed nothing from the electromagnetic theory,
we found that gravity and electricity share many properties in common
\textendash{} these similarities, in fact, are just a requirement
of special relativity that must apply to any physically acceptable
force field.   
}

\tableofcontents{}

\pagebreak{}

\section{The magnetism of gravity \textendash{} from Faraday to the present
days\label{Sec_Intro} }

\begin{figure}[H]
\centering\includegraphics[scale=0.75]{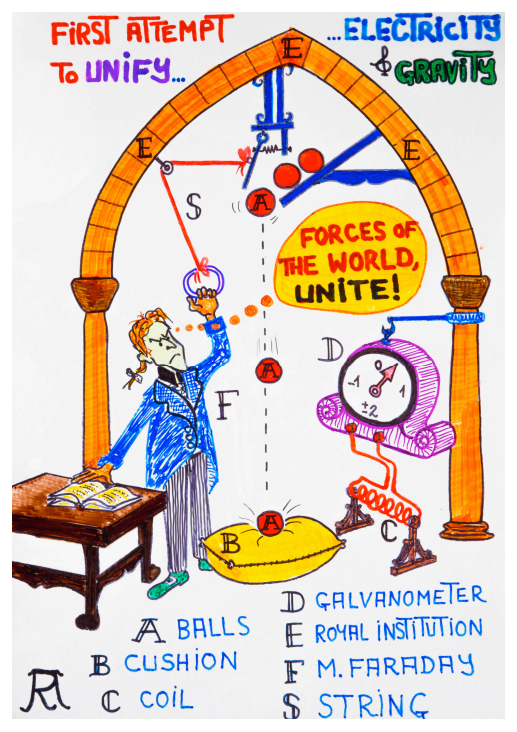}

\caption{Faraday attempts to detect the magnetism of gravity (illustration
reproduced with permission of A. de Rújula).}
\label{FigFaraday}
\end{figure}

Attempts to measure some kind of gravitational magnetism can be dated
back to the pioneer experiments of Faraday \cite{Faraday1843}, performed
about 1840. In these experiments, Faraday was wondering if gravity
would produce some effect analogous to the electromagnetic induction
that himself discovered before (see figure \ref{FigFaraday}). If
Faraday had been successful, the existence of gravitational magnetic
fields would have been established from the very beginning but, unfortunately,
he did not find any positive result. The first theoretical mention
of a possible analogy between electromagnetism and gravitation was
probably given by Maxwell in \cite{Maxwell1865}, although he did
not take that matter further. Some time later, Holzmüller \cite{Holzmuller1870}
and Tisserand \cite{Tisserand1872,Tisserand1890} independently proposed
that the Sun could exercise a gravitational magnetic force on the
surrounding planets (this was a naive attempt to explain the advance
of Mercury's perihelion through Weber's electrodynamics). It was only
at the end of the nineteenth century, however, that Heaviside presented
a direct and complete analogy between gravitation and electromagnetism
\cite{Heaviside1893}. Since then, the possibility of a gravitational
magnetism had been proposed by several authors, for instance, in the
first years of the special relativity, this theme appeared in the
works of Lorentz \cite{Lorentz1900}, Poincaré \cite{Poincare1906},
Minkowski \cite{Minkowski1910} and also Einstein \cite{Einstein1912}.

The way of looking at the magnetic effect of gravity changed, however,
after Einstein's formulation of general relativity and his theory
of gravitation \cite{Einstein1916}. In Einstein's theory, the gravitational
interaction is no longer regarded as a force, but only as an effect
of the spacetime curvature. By this reason, gravitation and electromagnetism
seemed to be very different theories and the earlier analogies between
these two force fields were almost despised. Actually, Einstein did
show in his seminal work that in the weak field regime Newton's theory
of gravitation can be recovered from a linearized form of his field
equations. However, he considered in that work only the case where
the speed of the gravitating bodies was very small when compared to
that of the light. If Einstein had allowed the bodies to move fast,
he probably would have rediscovered those aforementioned similarities
between gravitation and electromagnetism. Indeed, further developments
of the weak field method showed that gravity does present interesting
gravitomagnetic effects, for example, the frame-dragging effects of
de Sitter \cite{deSitter1916} and of Lense and Tirring \cite{Thirring1918,TirringLense1918}
as well as some effects analogous to the Thomas precession \cite{Thomas1927,Papapetrou1951}.

In modern times, the theoretical possibility of a magnetic gravitational
effect was considered from various perspectives. Some authors, notably
Jefimenko \cite{Jefimenko2000,Jefimenko2006}, suggested again a pure
analogy with electromagnetic theory. Benci and Fortunato \cite{Benci1998}
presented a mathematical argument (based on a variational principle)
that shows some connection between gravity and electromagnetism as
well. In the scope of special relativity, the existence of magnetic
gravitational fields were discussed by Salisbury \cite{Salisbury1974},
Lorrain \cite{Lorrain1975} and then Bedford \cite{Bedford1985} and
Kolbenstvedt \cite{Kolbenstvedt1988} (although these works discussed
only particular cases and different results were obtained, so that
no consensus was achieved regarding the exact formulæ describing the
gravitational magnetic field \cite{Khan1976,Fuchs1981}). Other more
recent considerations include the works of Behera and Naik \cite{BeheraNaik2004},
of Sattinger \cite{Sattinger2013} and also that of Ummarino and Gallerati
\cite{UmmarinoGallerati2017,Behera2017}, where the influence of gravitomagnetic
fields on superconductors were studied. It is nevertheless through
the general theory of relativity that the gravitomagnetic effects
have been extensively studied in the recent years. These methods are
commonly based on the linearized Einstein's field equations or other
higher-order expansions as the post-Newtonian approximations. These
approaches were developed by several authors, notably Jantzen, Bini,
Mashhoon, Iorio, among others $-$ see the reviews \cite{Jantzen1992,Mashhoon2000,Sanchez2001,Mashhoon2003,Tartaglia2003,Bini2008}
and references therein. Recently, exact methods have been also presented,
for instance, through the analysis of the tidal forces \cite{Costa2008,Costa2014}
or yet in the scope of gauge formulations of general relativity \cite{Clark2000,Ramos2010,Lasenby1998}.

Many experiments were performed as well with the aim of detecting
the magnetic effects of gravity. By the end of the fifties, Pugh \cite{Pugh1959}
and then Schiff \cite{Schiff1960A,Schiff1960B} already proposed a
way of measuring the gravitational frame-dragging effects with spatial
gyroscopes and, after soon, Forward \cite{Forward1963} speculated
about some gravitational devices analogues to electromagnetic machines;
no results however were found with these attempts either. The failure
of detecting any gravitational magnetic effect by these experiments
can be attributed to the very weakness of the gravitational interaction.
Currently, however, our technology has been greatly improved and we
are living an exciting moment where those gravitomagnetic effects
are beginning to be experimentally verified. For instance, scientists
behind \textsc{probe b} group have been conducted experiments that
confirm the existence of gravitomagnetic effects predicted by general
relativity, such as the geodetic and frame-dragging effects \cite{Ruggiero2002,Everitt2011,Iorio2011,Everitt2015}
and, very recently, we become aware of the astonishing news regarding
the first detection of gravitational waves, as reported by scientists
of \textsc{ligo} group \cite{Abbott2016,Abbott2016b,Abbott2017a,Abbott2017b}.
This was very promising for all physicists and we expect that many
other welcome surprises will appear in a near future. 

The purpose of this work is to show that the gravitomagnetic effects
can be also contemplated in the scope of special relativity. This
provides an alternative to those methods based on linearized Einstein
theory, which has the advantage of being simpler and that can be applied
whenever the gravitational field is weak, regardless of the speed
of the gravitating bodies. Furthermore, our approach is also very
enlightening, since it clarifies the nature and the origin of the
gravitomagnetic effects. Employing only the special theory of relativity,
we construct a fully covariant theory of gravitation w.r.t. the Lorentz
transformations. Although our theory is only an approximation of Einstein's
theory, it can be regarded as exact in a flat spacetime, in the same
sense as the theory of electromagnetism can be regarded as exact in
this case as well. In fact, starting uniquely from the special theory
of relativity (without using any result whatsoever of the electromagnetic
theory), we found that this covariant theory of gravitation shares
many properties in common with the electromagnetic theory. These similarities
are, although, not a matter of coincidence, but only a consequence
of the covariance of both theories regarding the Lorentz transformations.
Since this covariance is required for any physical theory, these properties
must hold to any acceptable relativistic force field as well.

This article is organized as follows: in section \ref{Sec_Exp} we
show from a thought experiment that gravity necessarily must have
a magnetic counterpart. In section \ref{Sec_Force} we derive the
exact formulæ describing the gravitomagnetic fields purely from covariance
requirements of special relativity theory. The question relying on
the invariance of the gravitational mass is analyzed in section \ref{Sec_Mass},
where we show that the gravitational mass should be regarded as an
invariant quantity, in the same footing as the electric charge. The
transformation laws for the gravitomagnetic fields w.r.t. inertial
reference frames are presented in section \ref{Sec_TransFields} and
in section \ref{sec_Maxwell} we prove that these fields satisfy a
set of differential equations which have the same form as the Maxwell
equations. The gravitomagnetic potentials are presented in section
\ref{Sec_GravPot}, where we argue that the vector potential can be
associated with the momentum of interaction between matter and fields
in the same manner as the scalar potential is associated with the
potential energy. The momentum and energy stored in the gravitational
fields are considered in section \ref{Sec_EPTensor} and a manifestly
covariant formulation of the theory is presented at section \ref{Sec_Covariant}.
The similarities and differences between gravity and electricity are
analyzed in section \ref{Sec-Diferences} and, finally, in section
\ref{Sec-Conclusions}, we present our conclusions and we discuss
some topics that can be exploited further (e.g., the possibility of
formulating a quantum version of this theory). Conventions, notations
and the vector identities used in this work are presented in the appendices.

\section{A thought experiment \label{Sec_Exp}}

We begin our approach with a thought experiment in order to show that
matter in motion must generate, besides the gravitational field, also
a magnetic field, which is in many ways analogous to the ordinary
magnetic field of electromagnetic theory.

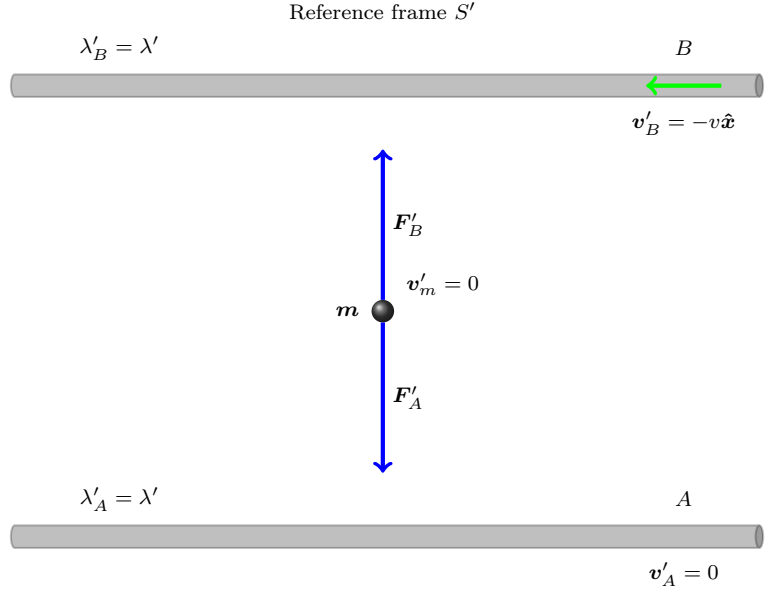
\begin{figure}
\tikzset{my cylinder/.style={cylinder, cylinder uses custom fill,rotate=0,outer sep=1pt}} 
\begin{centering}
\begin{tikzpicture}[thick]

\node [my cylinder, shape aspect=0.5,minimum width=0.3cm, minimum height=10cm, draw=black, cylinder end fill=darkgray!100, top color=black, bottom color=black, middle color=black, shading=axis, opacity=0.25] (cB) at (0,6){};

\node [my cylinder, shape aspect=0.5,minimum width=0.3cm, minimum height=10cm, draw=black,  cylinder end fill=darkgray!100, top color=black, bottom color=black, middle color=black, shading=axis, opacity=0.25] (cA) at (0,0){};

\shade [ball color=darkgray] (0,3) circle [radius=0.15cm];
\draw[->,ultra thick,blue]  (0,2.85) -- (0,0.85) node[midway, right, black] {$\bm{F}_A'$};
\draw[->,ultra thick,blue]  (0,3.15) -- (0,5.15) node[midway, right, black] {$\bm{F}_B'$};
\draw[->,ultra thick,green] (4.5,6) -- (3.5,6);

\coordinate (q1) at (-0.2,3);
\coordinate (q2) at (0.2,3.1);
\node[left] at (q1) {$\bm{m}$};
\node[above right] at (q2) {$\bm{v}'_m=0$};
\coordinate (Z) at (0,7); \node at (Z) {Reference frame $S'$};
\coordinate (A) at (4,0.5); \node at (A) {$A$};
\coordinate (B) at (4,6.5); \node at (B) {$B$};
\coordinate (C) at (-3.5,0.5); \node at (C) {$\lambda_A'=\lambda'$};
\coordinate (C) at (-3.5,6.5); \node at (C) {$\lambda_B' =\lambda'$};
\coordinate (E) at (4,5.5); \node at (E) {$\bm{v}'_B=-v\bm{\hat{x}}$};
\coordinate (f) at (4,-0.5); \node at (f) {$\bm{v}'_A=0$};
\end{tikzpicture} 
\par\end{centering}
\caption{The thought experiment in $S'$: Two parallel massive straight lines
exert a gravitational force in a particle of mass $m$, placed in
the midpoint of them. The densities of the lines are the same so that
the resulting force is null. The particle and line $A$ are at rest,
while line $B$ is moving with the velocity $\boldsymbol{v}=-v\hat{\boldsymbol{x}}$. }
\label{FigS'}
\end{figure}

This thought experiment can be described as follows: Suppose a system
composed of two parallel massive straight lines, say $A$ and $B$,
and also a particle of mass $m$ placed exactly at the midpoint of
these lines. For a reference frame $S'$, let those parallel lines
be in the $X'$ direction and suppose that both the particle and line
$A$ are at rest w.r.t. $S'$, while line $B$ is moving with a velocity
$-v$ in the same direction of the straight lines.

If the densities $\lambda_{A}'$ and $\lambda_{B}'$ of lines $A$
and $B$, respectively, are set to be the same in the reference frame
$S'$, that is, if we take $\lambda_{A}'=\lambda_{B}'=\lambda'$,
then the total gravitational force acting on the particle will be
zero. In fact, assuming the validity of Newton's law of gravitation
in $S'$, the force on the particle due to line $A$ will be given
by, 
\begin{equation}
\boldsymbol{F}'_{A}=-\frac{2gm\lambda'}{r'}\hat{\boldsymbol{y}},
\end{equation}
while the force due to line $B$ will be 
\begin{equation}
\boldsymbol{F}'_{B}=\frac{2gm\lambda'}{r'}\hat{\boldsymbol{y}},
\end{equation}
where $g$ is the Newton gravitational constant and $r'$ is the distance
between the particle and the lines. This is illustrated in the figure
\ref{FigS'}.

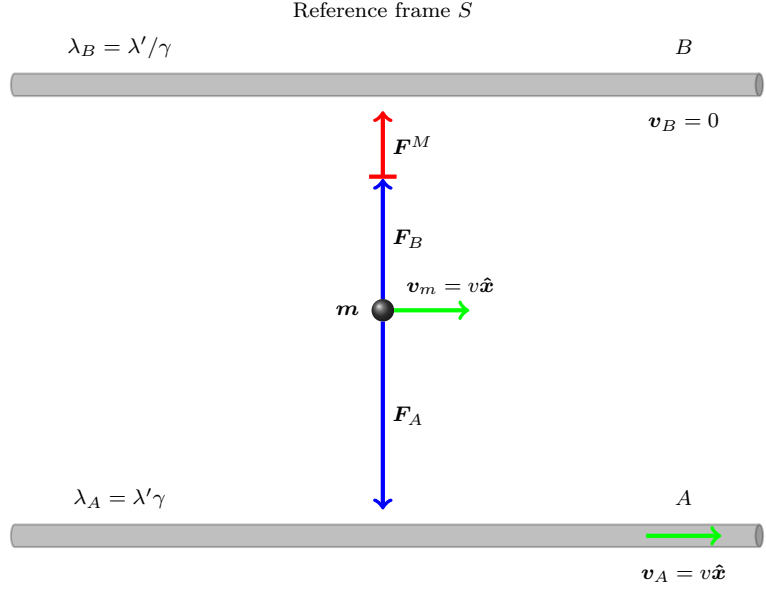
\begin{figure}[t]
\tikzset{my cylinder/.style={cylinder, cylinder uses custom fill,rotate=0,outer sep=1pt}} 
\begin{centering}
\begin{tikzpicture}[thick]

\node [my cylinder, shape aspect=0.5,minimum width=0.3cm, minimum height=10cm, draw=black, cylinder end fill=darkgray!100, top color=black, bottom color=black, middle color=black, shading=axis, opacity=0.25] (cB) at (0,6){};

\node [my cylinder, shape aspect=0.5,minimum width=0.3cm, minimum height=10cm, draw=black,  cylinder end fill=darkgray!100, top color=black, bottom color=black, middle color=black, shading=axis, opacity=0.25] (cA) at (0,0){};

\shade [ball color=darkgray] (0,3) circle [radius=0.15cm];
\draw[->,ultra thick,blue]    (0,2.85) -- (0,0.35) node[midway, right, black] {$\bm{F}_A$};
\draw[->,ultra thick,blue]   (0,3.15) -- (0,4.75) node[midway, right, black] {$\bm{F}_B$};
\draw[|->,ultra thick,red]     (0,4.75) -- (0,5.65) node[midway, right, black] {$\bm{F}^M$};
\draw[->,ultra thick,green]    (3.5,0) -- (4.5,0);
\draw[->,ultra thick,green] (0.15cm,3) -- (1.15,3);

\coordinate (q1) at (-0.2,3);
\coordinate (q2) at (0.2,3.1);
\node[left] at (q1) {$\bm{m}$};
\node[above right] at (q2) {$\bm{v}_m=v\bm{\hat{x}$}};
\coordinate (Z) at (0,7); \node at (Z) {Reference frame $S$};
\coordinate (A) at (4,0.5); \node at (A) {$A$};
\coordinate (B) at (4,6.5); \node at (B) {$B$};
\coordinate (C) at (-3.5,0.5); \node at (C) {$\lambda_A=\lambda'\gamma$};
\coordinate (C) at (-3.5,6.5); \node at (C) {$\lambda_B =\lambda'/\gamma$};
\coordinate (E) at (4,5.5); \node at (E) {$\bm{v}_B=0$};
\coordinate (f) at (4,-0.5); \node at (f) {$\bm{v}_A=v\bm{\hat{x}}$};
\end{tikzpicture} 
\par\end{centering}
\caption{The thought experiment in $S$: Here both the particle as line $A$
move with the velocity $\boldsymbol{v}=v\hat{\boldsymbol{x}}$, while
line $B$ is at rest. Thanks to the Lorentz contraction, the densities
of the lines are no longer the same, we get that $\lambda_{A}=\lambda'\gamma$
and $\lambda_{B}=\lambda'/\gamma$, so that the resulting gravitational
force exerted by the lines on the particle is not null anymore. The
principle of relativity, however, ensures that the total force acting
on the principle must also be null in $S$, which show us that there
must exist a gravitational magnetic force in order to restore that
symmetry. }
\label{FigS}
\end{figure}

Now, we might ask what happens when this system is observed from another
reference frame $S$, on which the particle and line $A$ have both
the velocity $\boldsymbol{v}=v\hat{\boldsymbol{x}}$, while line $B$
is now at rest. In the reference frame $S$, the mass densities $\lambda_{A}$
and $\lambda_{B}$ are no longer the same thanks to the Lorentz contraction
effect. In fact, since the mass of line $A$ is moving with velocity
$\boldsymbol{v}$, the mass density of this line, as measured by $S$,
is given by\footnote{An important question here is about the transformation law for the
gravitational mass density: Should the mass density $\lambda=m/V$
transform from $S'$ to $S$ as $\lambda=\gamma\lambda'$ or as $\lambda=\gamma^{2}\lambda'$?
If we regard $m$ as an invariant quantity, then it is clear that
the correct transformation law is $\lambda=\gamma\lambda'$, the $\gamma$-factor
coming from the transformation of the volume: $V=V'/\gamma$. However,
if we expressed the mass through Einstein's formula $E=mc^{2}$ then,
since the energy transforms as $E=\gamma E'$, we would be led to
the transformation law $\lambda=\gamma^{2}\lambda'$. Although may
appear at first that we have a paradoxical situation here, this is
not the case: as will be shown in more details in section \ref{Sec_Mass},
the correct transformation law is $\lambda=\gamma\lambda'$, so that
the mass must be regarded as an invariant quantity in the same manner
as the electric charge. In fact, the problem with the second possibility
is that mass density and energy density are actually two different
concepts \textendash{} the first is part of a spacetime vector while
the second is associated with a second-rank spacetime tensor \textendash ,
so the indiscriminate use of Einstein's formula above cannot be justified
(see \cite{Adler1987,Okun1989,Okun1989B,Okun2009,Jammer2000} where
this issue is analyzed in more details). \label{FootenoteM}} 
\begin{equation}
\lambda_{A}=\lambda'\gamma,
\end{equation}
while for line $B$, we have, since this line is now at rest, 
\begin{equation}
\lambda_{B}=\lambda'/\gamma.
\end{equation}
Therefore, in the reference frame $S$, the force that line $A$ exerts
on the particle is given by 
\begin{equation}
\boldsymbol{F}_{A}=-\gamma\frac{2gm\lambda'}{r}\hat{\boldsymbol{y}},
\end{equation}
while the force due to line $B$ is 
\begin{equation}
\boldsymbol{F}_{B}=\frac{1}{\gamma}\frac{2gm\lambda'}{r}\hat{\boldsymbol{y}}.
\end{equation}
Thus, the total gravitational force $\boldsymbol{F}=\boldsymbol{F}_{A}+\boldsymbol{F}_{B}$
acting on the particle would be 
\begin{equation}
\boldsymbol{F}=\frac{2gm\lambda'}{r}\left(\frac{1}{\gamma}-\gamma\right)\hat{\boldsymbol{y}}=-\frac{2gm\lambda'}{r}\left(\frac{\gamma v^{2}}{c^{2}}\right)\hat{\boldsymbol{y}},\label{FEGT}
\end{equation}
which, contrary to what is expected, is not null \textendash{} rather
it is directed towards line $A$. This odd result however cannot be
true, since the principle of relativity ensures that a particle cannot
fall towards line $A$ in a given reference frame while staying at
rest in another frame. Hence, we are led to the conclusion that, in
the reference frame $S$, there must exist some hidden force acting
on the particle, which must depend on its velocity, in order to balance
the gravitational forces. This force is nothing but the gravitational
analogue of the magnetic force. For the special case we considered
here, this magnetic force is given by 
\begin{equation}
\boldsymbol{F}^{M}=\frac{2gm\lambda'}{r}\left(\frac{\gamma v^{2}}{c^{2}}\right)\hat{\boldsymbol{y}}.\label{FMex1}
\end{equation}
In this way, we had shown from the principle of relativity that any
moving body necessarily must create a gravitational magnetic field.
This is illustrated in figure \ref{FigS}.

Let us present another argument to show that the force (\ref{FMex1})
can really be thought as a magnetic force associated with the gravity.
To this end, let us assume for the moment the existence \emph{a priori}
of a magnetic gravitational force $\boldsymbol{F}^{M}$ defined by
\begin{equation}
\boldsymbol{F}^{M}=m\boldsymbol{v}\times\boldsymbol{B},
\end{equation}
where $\boldsymbol{B}$ is the gravitational magnetic field, which
is supposed to satisfy a law analogous to Ampère's law of electromagnetic
theory, namely, 
\begin{equation}
\oint\boldsymbol{B}\cdot d\boldsymbol{s}=\mu\iint\boldsymbol{j}\cdot\mathrm{d}\boldsymbol{A},
\end{equation}
where $\boldsymbol{j}=\lambda\boldsymbol{v}/A$ is the mass current
density and $\mu$ is a constant yet to be determined. In the reference
frame $S$, the magnetic field generated by the system is only due
to line $A$, since line $B$ is at rest. Remembering that line $A$
moves with velocity $\boldsymbol{v}=v\hat{\boldsymbol{x}}$ in this
reference frame, and that its measured mass density should be $\lambda=\lambda'\gamma$,
so that $\boldsymbol{j}=\lambda'\gamma\boldsymbol{v}/A$, we have
from the Ampère's law above that 
\begin{equation}
\boldsymbol{B}=\frac{\mu\lambda'\gamma v}{2\pi r}\hat{\boldsymbol{z}}.
\end{equation}

Hence, the magnetic force acting on the particle (which moves with
the same velocity $\boldsymbol{v}=v\hat{\boldsymbol{x}}$) is 
\begin{equation}
\boldsymbol{F}^{M}=m\boldsymbol{v}\times\boldsymbol{B}=-\frac{\mu\lambda'm\gamma v^{2}}{2\pi r}\hat{\boldsymbol{y}}.\label{FMex2}
\end{equation}
Notice that this gravitational magnetic force is analogous to its
electromagnetic counterpart, except that it is repulsive rather than
attractive\footnote{In fact, if in our example we considered electric charges instead
of gravitational mass, the same result would be obtained, except that
the magnetic force would be attractive and more intense. This electromagnetic
case is explored, for instance, in Feynman's Lectures on Physics \cite{FeynmanLectures2013}.}. Moreover, we can see that, if we set 
\begin{equation}
\mu=-\frac{4\pi g}{c^{2}}=-\frac{\kappa}{2},
\end{equation}
where $\kappa$ is the Einstein gravitational constant, then both
expressions (\ref{FMex1}) and (\ref{FMex2}) become the same, which
show us that the force (\ref{FMex1}) can also be found without appeal
to the relativity theory, in the same fashion as the magnetic force
is calculated in the electromagnetic theory.

\section{ The magnetism of gravity as a consequence of Lorentz transformations.
The gravitomagnetic fields\label{Sec_Force} }

In the previous section we gave only a glimpse of the gravitomagnetic
effects provided by the special relativity theory. Now we shall begin
to develop a more rigorous theory of the gravitomagnetic theory. Except
when explicitly specified, hereafter we shall consider only the special
theory of relativity. This means that any effect due to the curvature
of spacetime implicated by the Einstein theory of gravitation will
be despised. In a sufficient distant region of the gravitational sources,
Newton's Law of gravitation gives a very good approximation for the
gravitational interaction, provided that the speed of the gravitating
bodies is small when compared with that of the light. Starting from
this, we can impose covariance of the gravitational force w.r.t. the
Lorentz transformations, from which exact formulæ for the gravitomagnetic
fields can be deduced. In this way, Newton's theory can be extended
to the case where the velocity of the orbiting bodies is comparable
to the light speed. We highlight again that no use of electromagnetic
theory (or any other analogy whatsoever) is employed in this approach
\textendash{} our results are derived purely from the transformation
laws of the special theory of relativity.

Therefore, let us take a given distribution of matter which is at
rest in a given inertial reference frame $S'$. This distribution
of matter creates a static gravitational field $\boldsymbol{G}'\left(x',y',z'\right)$,
so that a particle of mass $m$ situated at a point $\left(x',y',z'\right)$
of space will feel a gravitational force $\boldsymbol{F}'=\mathrm{d}\boldsymbol{p}'/\mathrm{d}t'$
given by 
\begin{equation}
\boldsymbol{F}'\left(x',y',z'\right)=m\boldsymbol{G}'\left(x',y',z'\right).\label{GField}
\end{equation}
We shall suppose further that this particle is moving with a velocity
$\boldsymbol{u}'$ w.r.t. $S'$, although the total force does not
depend on the particle velocity in $S'$.

Now we can ask what should be the (total) force $\boldsymbol{F}\left(x,y,z,t\right)$
acting on this particle w.r.t. another inertial reference frame $S$,
which moves with the velocity $\boldsymbol{v}=v_{x}\hat{\boldsymbol{x}}$
w.r.t. the frame $S'$. The answer for this question can be found
purely from the special theory of relativity $-$ it is not necessary
to know the nature of the force. In fact, the force $\boldsymbol{F}=\mathrm{d}\boldsymbol{p}/\mathrm{d}t$
measured in $S$ can be found directly through the respective transformation
formulæ provided by the special relativity theory \cite{Resnick1971,GriffithsE,Purcell},
which are, 
\begin{align}
F_{x}\left(x,y,z,t\right) & =F_{x}'\left(x',y',z'\right)+\frac{vu_{y}'}{c^{2}}\frac{F_{y}'\left(x',y',z'\right)}{1+u_{x}'v/c^{2}}+\frac{vu_{z}'}{c^{2}}\frac{F_{z}'\left(x',y',z'\right)}{1+u_{x}'v/c^{2}},\nonumber \\
F_{y}\left(x,y,z,t\right) & =\frac{1}{\gamma}\frac{F_{y}'\left(x',y',z'\right)}{1+u_{x}'v/c^{2}},\\
F_{z}\left(x,y,z,t\right) & =\frac{1}{\gamma}\frac{F_{z}'\left(x',y',z'\right)}{1+u_{x}'v/c^{2}}.\label{LTF}
\end{align}
Thus, from (\ref{GField}) we can find what is the total force $\boldsymbol{F}$
acting on the particle, as measured by the reference frame $S$. Notice,
however, that this force in general will depend on the time $t$,
even if the force $\boldsymbol{F}'$ does not depend on $t'$. This
is because the primed coordinates should still be eliminated through
the Lorentz transformations, 
\begin{equation}
t'=\gamma\left(t-\frac{v}{c^{2}}x\right),\qquad x'=\gamma\left(x-vt\right),\qquad y'=y,\qquad z'=z.\label{LT}
\end{equation}

From (\ref{LTF}) its not clear that the total force acting on the
particle is composed by gravitational and magnetic forces. Nevertheless
we can made this explicit if we write (\ref{LTF}) actually in terms
of the particle velocity $\boldsymbol{u}$, which is indeed the velocity
measured by the reference frame $S$. From the transformation laws
for the velocity components \cite{Resnick1971,GriffithsE,Purcell},
\begin{equation}
u_{x}=\frac{u_{x}'+v}{1+u_{x}'v/c^{2}},\qquad u_{y}=\frac{1}{\gamma}\frac{u_{y}'}{1+u_{x}'v/c^{2}},\qquad u_{z}=\frac{1}{\gamma}\frac{u_{z}'}{1+u_{x}'v/c^{2}},\label{LTV}
\end{equation}
we can eliminate $u_{x}'$, $u_{y}'$ and $u_{z}'$ in (\ref{LTF})
and, then, after simplify, the transformation laws for the force components
become, 
\begin{align}
F_{x}\left(x,y,z,t\right) & =F_{x}'(x',y',z')+\frac{\gamma v}{c^{2}}\left[u_{y}F_{y}'\left(x',y',z'\right)+u_{z}F_{z}'\left(x',y',z'\right)\right],\nonumber \\
F_{y}\left(x,y,z,t\right) & =\gamma\left(1-\frac{u_{x}v}{c^{2}}\right)F_{y}'\left(x',y',z'\right),\nonumber \\
F_{z}\left(x,y,z,t\right) & =\gamma\left(1-\frac{u_{x}v}{c^{2}}\right)F_{z}'\left(x',y',z'\right).\label{LTF1}
\end{align}

These formulæ are the key point of our approach. From this we can
plainly see that the total force acting on the particle can be split
into two parts: a part which does not depend on the particle velocity
$\boldsymbol{u}$ and a part that does depend. In this way, we define
the (electric-like) \emph{gravitational force} acting on the particle
as that part of the \emph{gravitomagnetic} force which does not depend
on the velocity of the particle and, on the other hand, we define
the (gravitational) \emph{magnetic force} as that part of the gravitomagnetic
force which does depend on the particle velocity. In terms of components,
we get, for the gravitational force, 
\begin{equation}
F_{x}^{G}\left(x,y,z,t\right)=F_{x}'\left(x',y',z'\right),\qquad F_{y}^{G}\left(x,y,z,t\right)=\gamma F_{y}'\left(x',y',z'\right),\qquad F_{z}^{G}\left(x,y,z,t\right)=\gamma F_{z}'\left(x',y',z'\right),\label{FG}
\end{equation}
while for the components of the magnetic force, we get, 
\begin{align}
F_{x}^{M}\left(x,y,z,t\right) & =+\frac{\gamma v}{c^{2}}\left[u_{y}F_{y}'\left(x',y',z'\right)+u_{z}F_{z}'\left(x',y',z'\right)\right],\nonumber \\
F_{y}^{M}\left(x,y,z,t\right) & =-\frac{\gamma v}{c^{2}}\left[u_{x}F_{y}'\left(x',y',z'\right)\right],\nonumber \\
F_{z}^{M}\left(x,y,z,t\right) & =-\frac{\gamma v}{c^{2}}\left[u_{x}F_{z}'\left(x',y',z'\right)\right].\label{FM}
\end{align}
It should be kept in mind, although, that the Lorentz transformations
should still be used in order to express those quantities in terms
of $x$, $y$, $z$ and $t$.

Now, it is straightforward to show that the total force $\boldsymbol{F}=\boldsymbol{F}^{G}+\boldsymbol{F}^{M}$
can be expressed in the Lorentz form, 
\begin{equation}
\boldsymbol{F}\left(x,y,z,t\right)=m\left[\boldsymbol{G}\left(x,y,z,t\right)+\boldsymbol{u}\times\boldsymbol{B}\left(x,y,z,t\right)\right],\label{LorentzForce}
\end{equation}
by introducing the gravitational field 
\begin{equation}
\boldsymbol{G}\left(x,y,z,t\right)=\boldsymbol{F}^{G}\left(x,y,z,t\right)/m,\label{FieldE}
\end{equation}
and defining the magnetic field as 
\begin{equation}
\boldsymbol{B}\left(x,y,z,t\right)=\frac{\boldsymbol{v}}{c^{2}}\times\boldsymbol{G}\left(x,y,z,t\right).\label{FieldB}
\end{equation}

The formulæ deduced above are quite general: they hold to any distribution
of mass which is at rest in a given reference frame $S'$ and, consequently,
moves with a velocity $\boldsymbol{v}$ in another reference frame
$S$. Besides, since the total force acting on the particle given
by the Lorentz formula (\ref{LorentzForce}) makes no direct mention
to what reference frame it refers (the force depends only on the gravitomagnetic
fields and the instantaneous velocity of the particle), this means
that the Lorentz force should be valid in any inertial reference frame.
This result will be confirmed in section \ref{Sec_TransFields}.

Let us apply now the arguments presented above\footnote{We highlight that this approach had been recently employed to explain
the \emph{Supplee submarine paradox} \cite{Supplee1989,Matsas2003}
in the framework of special relativity in \cite{Vieira2017}.} in the simplest case of a point-like particle of mass $M$, which
we suppose to be at rest in the origin of the reference frame $S'$
at $t'=0$. The gravitational field generated by this particle is
given, in the reference frame $S'$, by Newton's law of gravitation:
\begin{equation}
\boldsymbol{G}'\left(x',y',z'\right)=-gM\frac{x'\hat{\boldsymbol{x}}+y'\hat{\boldsymbol{y}}+z'\hat{\boldsymbol{z}}}{\left(x^{\prime2}+y^{\prime2}+z^{\prime2}\right)^{3/2}}.\label{Newton}
\end{equation}
Equations (\ref{FieldE}) and (\ref{FieldB}) give respectively the
gravitational and magnetic fields in the frame $S$, namely, 
\begin{equation}
\boldsymbol{G}\left(x,y,z,t\right)=-\gamma\frac{gM}{R^{3}}\left[\left(x-vt\right)\hat{\boldsymbol{x}}+y\hat{\boldsymbol{y}}+z\hat{\boldsymbol{z}}\right],
\end{equation}
and 
\begin{equation}
\boldsymbol{B}\left(x,y,z,t\right)=\frac{\gamma v}{c^{2}}\frac{gM}{R^{3}}\left(z\hat{\boldsymbol{y}}-y\hat{\boldsymbol{z}}\right)=-\frac{\gamma}{c^{2}}\frac{gM}{R^{3}}\left(\boldsymbol{v}\times\boldsymbol{R}\right),
\end{equation}
where $\boldsymbol{R}=\gamma\left(x-vt\right)\hat{\boldsymbol{x}}+y\hat{\boldsymbol{y}}+z\hat{\boldsymbol{z}}$
is the vector that measures the distance $R=\left|\boldsymbol{R}\right|$
from the present position of the particle (\emph{i.e.}, at $t=0$)
to the space point where the fields are evaluated; we had also made
use of the Lorentz transformations (\ref{LT}) to eliminate the coordinates
$x'$, $y'$ and $z'$. Notice further that the formula for the magnetic
field is analogous to the relativistic Biot-Savart law of electromagnetism.

Finally, from the Lorentz formula (\ref{LorentzForce}) we find the
gravitational and the magnetic forces acting on the particle: 
\begin{equation}
\boldsymbol{F}^{G}\left(x,y,z,t\right)=-\gamma\frac{gMm}{R^{3}}\left[\left(x-vt\right)\hat{\boldsymbol{x}}+y\hat{\boldsymbol{y}}+z\hat{\boldsymbol{z}}\right],
\end{equation}
and 
\begin{equation}
\boldsymbol{F}^{M}\left(x,y,z,t\right)=-\frac{\gamma v}{c^{2}}\frac{gMm}{R^{3}}\left[\left(u_{y}y+u_{z}z\right)\hat{\boldsymbol{x}}-u_{x}y\hat{\boldsymbol{y}}-u_{x}z\hat{\boldsymbol{z}}\right],
\end{equation}
so that the total force is given by 
\begin{equation}
\boldsymbol{F}\left(x,y,z,t\right)=-\gamma\frac{gMm}{R^{3}}\left\{ \left[\left(x-vt\right)+\frac{v}{c^{2}}\left(u_{y}y+u_{z}z\right)\right]\hat{\boldsymbol{x}}+\left(1-\frac{vu_{x}}{c^{2}}\right)\left(y\hat{\boldsymbol{y}}+z\hat{\boldsymbol{z}}\right)\right\} .
\end{equation}

\section{Comparison with the linearized theory of general relativity}

In the previous section we derived the formulæ describing the gravitomagnetic
fields according to the special theory of relativity. The standard
way in which these phenomena are studied, however, is through the
general theory of relativity, usually considering the linearized or
post-Newtonian approximations of Einstein's gravitation theory. Our
purpose in this section is to briefly comment on the differences regarding
the description of the gravitomagnetic phenomena implicated by these
two approaches.

In the simplest case of the linearized Einstein theory, we consider
that spacetime is almost flat, so that the metric $\xi^{\mu\nu}$
of the curved spacetime can be split as $\xi^{\mu\nu}=\eta^{\mu\nu}+\zeta^{\mu\nu}$,
where $\eta^{\mu\nu}$ is the Minkowski metric of the flat spacetime
and $\zeta^{\mu\nu}$ is a small perturbation. With these assumptions,
the gravitomagnetic phenomena are deduced as follows: we first need
to solve the Einstein field equations, taking into account the approximation
employed (i.e., disregarding all quantities which are of second, or
higher, order in $\zeta^{\mu\nu}$). Then, the geodesic equations
should be solved, so that we can obtain the acceleration of a particle
that moves only under the influence of the gravitational fields. Finally,
after we know the acceleration of the particle, then we can get the
gravitomagnetic forces acting on it \cite{Misner1973,Landau1980,Pauli1981,Wald1984}.

In this way, if we assume that the source of the gravitational field
is at rest and the orbiting particle does not move so fast, then we
are led to Newton's theory of gravitation \cite{Einstein1916}. On
the other hand, in order to obtain gravitomagnetic effects, we should
allow the source to move. In this case, the linearized Einstein field
equations can be reduced, in certain circumstances, to a set of equations
resembling the Maxwell equations of electromagnetism, although they
are not the same. For instance, solving the geodesic equations we
find that the gravitomagnetic force is given by a modified Lorentz
formula \cite{Wald1984}, namely, 
\begin{equation}
\boldsymbol{F}=m\left(\boldsymbol{G}+4\boldsymbol{u}\times\boldsymbol{B}\right).\label{LorentzGR}
\end{equation}
Therefore, we can see that the linearized theory of general relativity
predicts different results if compared with our theory. This is expected,
of course, since we did not take into account the effects of spacetime
curvature in our approach. In fact, the apparition of this factor
of $4$ in the magnetic force above seems to be of the same nature
as that factor of $2$ which was obtained by Einstein in his calculation
of the gravitational bending of light, after he took into account
the curvature of spacetime \cite{Einstein1916}. 

Besides, it should be noticed that there are some implicit and subtle
assumptions considered in the approximation above. First of all, when
solving the geodesic equations we should actually go beyond the first
order approximation because, if we did not, the geodesics would become
the same as that obtained in a flat spacetime \cite{Wald1984}. Hence,
terms of different orders should be carefully kept in order to compare
the resulting equations with the electromagnetic theory. Secondly,
these results are generally obtained only when specific gauge conditions
are imposed to the field equations \textendash{} the theory thus obtained
is not gauge-invariant in the same sense as the electromagnetic theory
is, neither it is covariant regarding the Lorentz transformations.
Moreover, we should remember that Einstein's field equations actually
contain terms which depend on the pressure and stresses as well and,
if these terms were taken into account, then we would also get effects
with no electromagnetic analogue $-$ see, for instance, the Matte-Bel
decomposition of Riemann tensor \cite{Matte1953,Bel1958}. It should
be mentioned further that attempts to remove the factor 4 in (\ref{LorentzGR})
were also considered \cite{Jantzen1992,Mashhoon2000,Sanchez2001,Mashhoon2003,Tartaglia2003,Bini2008};
these attempts however basically try to redefine the gravitomagnetic
fields or potentials but, in general, the resulting gravitational
Maxwell equations become different from that obtained in the electromagnetic
theory (see also \cite{Bouda2010,Bakopoulos2014,Bakopoulos2016} for
alternative perturbative methods that seems to successfully reproduce
the Maxwell equations for gravity without the factor of 4 in the magnetic
force above).

From what has been said, we can see that our approach is quite different
from that obtained through weak field approximations of general relativity.
The gravitational magnetic field obtained here is due only to the
motion of matter (without restriction to the velocity of the gravitational
source), while in the general relativity framework it is mainly due
to small perturbations of the metric. Therefore the concepts involved
in both approaches are quite different in nature and we cannot expected
that the results from both theories exactly agree. In fact, the results
that will be discussed in the present work should be regarded as a
first approximation to the theory of gravitomagnetic phenomena, valid
with good precision at great distances of the gravitational sources
so that the gravitation interaction can be described by an effective
force field; the gravitomagnetic effects predicted by the linearized
theory of general relativity should be regarded as the next more accurate
approximation, namely, that one on which the spacetime is supposed
to be just a little bit curved.

\section{Transformation laws for the mass and current densities and the principle
of equivalence\label{Sec_Mass}}

An important issue that is very debated in discussions about the foundations
of relativity theory, and that is very important in our theory as
well, is the question of invariance of the gravitational mass w.r.t.
the Lorentz transformations. It is unanimous accepted fact that electric
charge is such an invariant, but no agreement at all is reached for
the gravitational mass. Our intention in this section is to clarify
this matter.

The beginning of this controversy can be dated to the first formulations
of the relativity theory and it also relies on the definition of mass
that is employed \cite{Adler1987,Okun1989,Okun1989B,Okun2009,Jammer2000}.
First of all, we have at least two different conceptions of mass:
the \emph{inertial mass} and the \emph{gravitational mass}. In the
old times of special relativity, several authors (Einstein including)
have considered that the inertial mass depends on the velocity through
the equation $m(u)=m_{0}\gamma_{u}$, where $m_{0}$ is the so-called
\emph{rest mass}. The development of relativity showed us, however,
that it is more reasonable to regard the inertial mass as an invariant
quantity, from which the theory of relativity can be formulated in
a simpler and more elegant way\footnote{Notice that the dependence of the mass with the velocity can be seen
either as a matter of convention or as a true experimental issue.
The second interpretation, however, face difficulties because it is
hard to conceive an independent measurement of the inertial mass,
since it is generally attached to other physical quantities as the
momentum or energy, for instance.}. In the case of the gravitational mass this issue becomes most dramatic
because it seems, at first sight, that if the gravitational mass were
dependent on the velocity, then the strength of the gravitational
field of moving bodies would be greater than that of bodies at rest.
This, however, is not the case, since we had shown that the total
force acting on the moving particle is provided by the transformations
(\ref{LTF1}) and the expression for the gravitational force in the
proper frame of the source. Hence, no matter what definition of inertial
mass we use: a different definition would just lead to a different
definition of the gravitomagnetic fields $-$ the forces cannot be
changed by a mere redefinition of the mass.

Independently on what definition of mass is used, what is important
to be noticed is that the source of the gravitational interaction
must be regarded as an invariant quantity\footnote{This should be true both in the special as in the general theory of
relativity. As it is known, the gravitational interaction in Einstein's
theory is determined by the spacetime momentum-flux tensor $T^{\mu\nu}$.
However, the \emph{strength} of the gravitational interaction (\emph{i.e.},
the strength of the spacetime curvature) must be determined only by
the \emph{proper }spacetime momentum-flux tensor, otherwise a single
particle could even become a black-hole in a reference frame where
it has a great enough speed, which is clearly an absurd. Therefore,
the effects implicated by the motion of matter should be of the same
nature as the magnetic effects introduced here, although its mathematical
description becomes more complex due to the non-linearity of Einstein's
equations. \label{FootnoteT}}, namely, the mass (i.e., the rest mass, if the older interpretation
is adopted), so that the gravitational mass density must transform,
from the reference frame $S'$ to $S$ as, 
\begin{equation}
\rho=\gamma\rho'.\label{MassDensity}
\end{equation}
As commented in the footnote \ref{FootenoteM}, a naive argument would
suggest that the gravitational mass density should transform as $\rho=\gamma^{2}\rho'$
instead, but we shall see in the sequel that this claim is wrong.
In fact, the validity of (\ref{MassDensity}) is ultimately a consequence
of the equivalence principle. To see that, notice that the mass of
a particle can be determined by the relativistic relation,
\begin{equation}
m^{2}c^{4}=E^{2}-c^{2}p^{2},
\end{equation}
where $E$ is the particle energy and $\boldsymbol{p}$ its the momentum.
Let us to consider, then, the following example: suppose an ensemble
of particles distributed in a uniform way and at rest in the reference
frame $S'$. Let $\sigma'=N'/V'$ be the concentration of particles
(i.e., the number of particles per unit volume), as measured by $S'$.
Now we ask what will be this concentration $\sigma=N/V$ w.r.t. the
reference frame $S$, where the ensemble moves with the velocity $\boldsymbol{v}=v\hat{\boldsymbol{x}}$.
It is clear that this concentration is given by $\sigma=\gamma\sigma'$,
since the number of particles per unit volume will be increased by
a $\gamma$-factor, thanks to the Lorentz contraction. Now, if we
consider that each particle of the ensemble has an inertial mass $\mu$,
so that the total mass contained in a giving volume is $m$, then
it follows from the fact that the inertial mass is an invariant of
Lorentz that the inertial mass density must transform according to
(\ref{MassDensity}). But since the inertial mass should equal the
gravitational mass if the equivalence principle is true, we may conclude
that the same transformation law (\ref{MassDensity}) must hold for
the gravitational mass density. Besides, since the Lorentz formula
should be valid in any inertial reference frame (see section \ref{Sec_TransFields}),
the invariance of the gravitational mass also follows, and further,
as we shall see in section \ref{sec_Maxwell}, this is also necessary
for the gravitational Maxwell equations be covariant w.r.t. the Lorentz
transformations \textendash{} in short, the invariance of the inertial
and gravitational mass is mandatory for a covariant theory.

At this point, we would like to remark that the concepts of mass and
current densities are not the same as the concepts of energy and momentum
densities $-$ this is the source of many confusions in the literature,
which usually leads to mistakes as that one commented above. In fact,
since in relativity theory energy and momentum are regarded as the
components of a spacetime vector, the same cannot be true for their
densities. Indeed, from the transformation law for the energy and
momentum from $S'$ to $S,$ 
\begin{equation}
E=\gamma E',\qquad\boldsymbol{p}=\frac{\gamma\boldsymbol{v}}{c^{2}}E',
\end{equation}
it follows that the energy density $\varepsilon$ and also the momentum
density $\boldsymbol{\pi}$ must transform as 
\begin{equation}
\varepsilon=\gamma^{2}\varepsilon',\qquad\boldsymbol{\pi}=\frac{\gamma^{2}\boldsymbol{v}}{c^{2}}\varepsilon',
\end{equation}
if we neglect the stresses contributions \textendash{} so we can see
where the confusion comes from. The densities of energy and momentum
do not form a spacetime vector; on the contrary, they are (together
with the stresses) the components of a two-rank tensor $-$ the \emph{spacetime
momentum-flux tensor}. See sections \ref{Sec_EPTensor} and \ref{Sec_Covariant}
for more details.

Notice moreover that we can define in $S$ the \emph{current density}
associated with the motion of the ensemble as 
\begin{equation}
\boldsymbol{j}=\boldsymbol{v}\rho.
\end{equation}
The conservation law for the number of particles leads us then to
the conservation of matter. This means that the mass and current densities
satisfy the \emph{continuity equation}, 
\begin{equation}
\partial_{t}\rho+\boldsymbol{\nabla}\cdot\boldsymbol{j}=0.\label{Continuity}
\end{equation}
This equation, as a matter of fact, also results directly from covariance
requirements: take a static distribution of matter in $S'$, so that
$\mathrm{d}\rho'/\mathrm{d}t'=0.$ Since $\rho'=\rho/\gamma$, we
get that 
\begin{equation}
\frac{\mathrm{d}\rho'}{\mathrm{d}t'}=\gamma\left(\partial_{t}\rho'+v\partial_{x}\rho'\right)=\partial_{t}\rho+v\partial_{x}\rho=0,\label{ContX}
\end{equation}
where we had used the relation $\mathrm{d}/\mathrm{d}t'=\gamma\left(\partial/\partial t+v\partial/\partial x\right)$,
deduced from the Lorentz transformations. Thus, since in the case
considered here we have $\boldsymbol{j}=v\rho\hat{\boldsymbol{x}}$,
this means that (\ref{ContX}) is equivalent to the continuity equation
(\ref{Continuity}).

General formulæ for the transformations of the mass and current densities
can be found by considering the reference frames $S$ and $S''$.
In the example above, suppose that the ensemble moves with the velocity
$\boldsymbol{u}$ w.r.t. the reference frame $S$ and with the velocity
$\boldsymbol{u}''$ w.r.t. $S''$, and also consider that the particles
composing the ensemble are still uniformly distributed. Therefore,
the mass and current densities will be $\rho$ and $\boldsymbol{j}$
when measured by $S$ and $\rho''$ and $\boldsymbol{j}''$ w.r.t.
$S''$. Then it is easy to show from the Lorentz transformations for
the velocity components (\ref{LTV}) that the mass density and the
components of the current densities transform as 
\begin{equation}
\rho''=\gamma\left(\rho-vj_{x}\right),\qquad j_{x}''=\gamma\left(j_{x}-\frac{v}{c^{2}}\rho\right),\qquad j_{y}''=j_{y},\qquad j_{z}''=j_{z},\label{MCT}
\end{equation}
that is, the mass density and the components of the current density
form a spacetime vector $-$ the \emph{spacetime current density}.

These results can also be generalized to the case where the mass and
current densities are not uniform nor constant in time. To this end,
we may consider that the number of particles of the ensemble goes
to infinity, while the distance between them goes to zero (so that
we get a continuous distribution of matter) and also that those particles
contained in a given volume element have a specific velocity. Then,
all we need to do is repeat the arguments above to each one of these
volume elements. Notice, however, that each volume element will be
led to a different time when the Lorentz transformation is performed.
It might appear, therefore, that these transformation laws for the
mass density $\rho(x,y,z,t)$ and the current density $\boldsymbol{j}(x,y,z,t)$
cannot be defined to extended objects. The reason is that, if these
quantities are measured in $S$ in the same instant of time, say $t_{0}$,
then the mass density $\rho''(x'',y'',z'',t'')$ and the current density
$\boldsymbol{j}''(x'',y'',z'',t'')$ will be led to volume elements
\emph{in different instants of time}, but when a given observer in
$S''$ measures these densities, he always do it \emph{in the  same
instant of time} $t_{0}''$, therefore it is not clear at first that
he would get the same result as given by (\ref{MCT}). Fortunately,
he indeed gets the same result, as was proven by Podolsky \cite{Podolsky1947}
for the analogous electromagnetic case.

\section{Transformation laws for the fields\label{Sec_TransFields}}

In section \ref{Sec_Force} we have shown that any massive moving
body must present both a gravitational as a magnetic field. Now in
this section we shall deduce the transformation laws for these gravitomagnetic
fields from an inertial reference frame to another.

To this end, consider that in the reference frame $S$ there exists
a moving distribution of matter which creates both a gravitational
field $\boldsymbol{G}\left(x,y,z,t\right)$ as a magnetic field $\boldsymbol{B}\left(x,y,z,t\right)$.
If a particle of mass $m$ moves in this gravitomagnetic field, say
with an instantaneous velocity $\boldsymbol{u}$ measured at $t=0$,
then it will be subject to a gravitomagnetic force that is given by
the Lorentz formula (\ref{LorentzForce}), as we showed in section
\ref{Sec_Force}. In terms of its components, the total force acting
on the particle will be, therefore, 
\begin{equation}
F_{x}=mG_{x}+m\left(u_{y}B_{z}-u_{z}B_{y}\right),\qquad F_{y}=mG_{y}+m\left(u_{z}B_{x}-u_{x}B_{z}\right),\qquad F_{z}=mG_{z}+m\left(u_{x}B_{y}-u_{y}B_{x}\right).\label{LorentzComp}
\end{equation}

Now we may ask what will be the total force acting on this particle
for another reference frame $S''$, where the (still undetermined)
gravitomagnetic fields are $\boldsymbol{G}''\left(x'',y'',z'',t''\right)$
and $\boldsymbol{B}''\left(x'',y'',z'',t''\right)$, respectively,
and where the particle moves with the instantaneous velocity $\boldsymbol{u}''$.
The expressions for the force components in $S''$ can be found through
the inverse of the force transformation laws (\ref{LTF1}), namely,
\begin{equation}
F_{x}''=F_{x}-\frac{\gamma v}{c^{2}}\left(u_{y}''F_{y}+u_{z}''F_{z}\right),\qquad F_{y}''=\gamma\left(1+\frac{u_{x}''v}{c^{2}}\right)F_{y},\qquad F_{z}''=\gamma\left(1+\frac{u_{x}''v}{c^{2}}\right)F_{z},
\end{equation}
where we omitted dependence of the forces on $x$, $y,$ $z$, $t$
etc. for short. In fact, replacing the components $F_{x}$, $F_{y}$
and $F_{z}$ in this equation by their respective expressions given
by (\ref{LorentzComp}) and using (\ref{LTV}) to eliminate the velocities
$u_{x}$, $u_{y}$ and $u_{z}$, we get, after simplification, the
expressions, 
\begin{align}
F_{x}'' & =mG_{x}+\gamma mu_{y}''\left(B_{z}-\frac{v}{c^{2}}G_{y}\right)+\gamma mu_{z}''\left(B_{y}+\frac{v}{c^{2}}G_{z}\right),\nonumber \\
F_{y}'' & =\gamma m\left(G_{y}-vB_{z}\right)+mu_{z}''B_{x}-\gamma mu_{x}''\left(B_{z}-\frac{v}{c^{2}}G_{y}\right),\nonumber \\
F_{z}'' & =\gamma m\left(G_{z}+vB_{y}\right)+\gamma mu_{x}''\left(B_{y}+\frac{v}{c^{2}}G_{z}\right)-mu_{y}''B_{x}.
\end{align}

Now, we may realize that this force can also be written as a gravitational
force plus a magnetic force. In fact, the gravitational force is defined
as the part of the total force that does not depend on the particle
velocity $\boldsymbol{u}''$, and it is given by,
\begin{equation}
F_{x}^{G\prime\prime}=mG_{x},\qquad F_{y}^{G\prime\prime}=m\gamma\left(G_{y}-vB_{z}\right),\qquad F_{z}^{G\prime\prime}=m\gamma\left(G_{z}+vB_{y}\right),
\end{equation}
while the magnetic force is defined as the part that does depend on
the particle velocity and, hence, it is given by 
\begin{align}
F_{x}^{B\prime\prime} & =m\left[u_{y}''\gamma\left(B_{z}-\frac{v}{c^{2}}G_{y}\right)+u_{z}''\gamma\left(B_{y}+\frac{v}{c^{2}}G_{z}\right)\right],\nonumber \\
F_{y}^{B\prime\prime} & =m\left[u_{z}''B_{x}-u_{x}''\gamma\left(B_{z}-\frac{v}{c^{2}}G_{y}\right)\right],\nonumber \\
F_{z}^{B\prime\prime} & =m\left[u_{x}''\gamma\left(B_{y}+\frac{v}{c^{2}}G_{z}\right)-u_{y}''B_{x}\right].
\end{align}
Therefore, we conclude that in the reference frame $S''$ the total
force can also be written in the Lorentz form as 
\begin{equation}
\boldsymbol{F}''=m\left(\boldsymbol{G}''+\boldsymbol{u}''\times\boldsymbol{B}''\right),
\end{equation}
provided that the gravitomagnetic fields $\boldsymbol{G}''$ and $\boldsymbol{B}''$,
as measured by $S''$, are related to the gravitomagnetic fields $\boldsymbol{G}$
and $\boldsymbol{B}$, as measured by $S$, by the formulæ, 
\begin{equation}
G_{x}''=G_{x},\qquad G_{y}''=\gamma\left(G_{y}-vB_{z}\right),\qquad G_{z}''=\gamma\left(G_{z}+vB_{y}\right),\label{GLT}
\end{equation}
and 
\begin{equation}
B_{x}''=B_{x},\qquad B_{y}''=\gamma\left(B_{y}+\frac{v}{c^{2}}G_{z}\right),\qquad B_{z}''=\gamma\left(B_{z}-\frac{v}{c^{2}}G_{y}\right).\label{BLT}
\end{equation}

In this way we have proved that the Lorentz force holds for any inertial
reference frame, a result that was indeed anticipated at section \ref{Sec_Force}.
At the same time, we have found the transformation laws for the gravitomagnetic
fields, expressed through (\ref{GLT}) and (\ref{BLT}).

\section{The gravitomagnetic Maxwell equations\label{sec_Maxwell}}

In this section we shall concern ourselves with the differential equations
governing the gravitomagnetic fields. As we have seen, a static distribution
of mass (for instance, w.r.t. the reference frame $S'$) generates
only a static gravitational field $\boldsymbol{G}'$. In the approximation
considered here, where the gravitational force can be interpreted
as an ordinary force field, the universal Newton's law of gravitation
holds, from which follows that the gravitational field $\boldsymbol{G}'$
enjoys the following properties:
\begin{enumerate}
\item $\boldsymbol{G}'$ satisfies the Gauss law, 
\begin{equation}
\boldsymbol{\nabla}'\cdot\boldsymbol{G}'=-4\pi g\rho';\label{DivS'}
\end{equation}
\item $\boldsymbol{G}'$ is an irrotational field, 
\begin{equation}
\boldsymbol{\nabla}'\times\boldsymbol{G}'=0;\label{RotG'}
\end{equation}
\item $\boldsymbol{G}'$ is independent of time, 
\begin{equation}
\partial_{t}'\boldsymbol{G}'=0.\label{dGdt'}
\end{equation}
\end{enumerate}
From these properties and the Lorentz transformations we can find
what should be the differential equations describing the gravitomagnetic
fields in the reference frame $S$. For this end, remember that in
$S$ there exist both a gravitational field $\boldsymbol{G}$ as a
magnetic field $\boldsymbol{B}$ and that the magnetic field is related
to the gravitational field by formula (\ref{FieldB}), that is, $\boldsymbol{B}=\boldsymbol{v}/c^{2}\times\boldsymbol{G}$.
Hence, to find the differential equations satisfied by the gravitomagnetic
fields, we need to express the derivatives $\partial_{x}'$, $\partial_{y}'$,
$\partial_{z}'$ and $\partial_{t}'$ in terms of $\partial_{x}$,
$\partial_{y}$, $\partial_{z}$ and $\partial_{t}$. We easily get
the relations necessary: 
\begin{equation}
\partial_{x}'=\gamma\left(\partial_{x}+\frac{v}{c^{2}}\partial_{t}\right),\qquad\partial_{y}'=\partial_{y},\qquad\partial_{z}'=\partial_{z},\qquad\partial_{t}'=\gamma\left(\partial_{t}+v\partial_{x}\right).\label{Derivs}
\end{equation}
We shall need further the transformations formulæ for the gravitational
field given at (\ref{GLT}), which in this case reduce to the equations
\begin{equation}
G_{x}'=G_{x},\qquad G_{y}'=G_{y}/\gamma,\qquad G_{z}'=G_{z}/\gamma,\label{Gprimed}
\end{equation}
since the magnetic field is null in $S'$. Now, inserting the relations
(\ref{Derivs}) and (\ref{Gprimed}) in (\ref{dGdt'}), we get, 
\begin{equation}
\partial_{x}G_{x}+v\partial_{t}G_{x}=0,\qquad\partial_{x}G_{y}+v\partial_{t}G_{y}=0,\qquad\partial_{x}G_{z}+v\partial_{t}G_{z}=0,\label{dGdT}
\end{equation}
which will be of importance in the what follows.

Therefore, let us begin analyzing what should be the divergence of
the gravitational field in the reference frame $S$. Replacing the
coordinate derivatives present in (\ref{DivS'}) by the relations
(\ref{Derivs}) we get, at once, 
\begin{align}
\partial_{x}'G_{x}'+\partial_{y}'G_{y}'+\partial_{z}'G_{z}' & =-4\pi g\rho',\nonumber \\
\gamma\left(\partial_{x}+\frac{v}{c^{2}}\partial_{t}\right)G_{x}+\partial_{y}\left(\frac{G_{y}}{\gamma}\right)+\partial_{z}\left(\frac{G_{z}}{\gamma}\right) & =-4\pi g\rho',\nonumber \\
\gamma\left(1-\frac{v^{2}}{c^{2}}\right)\partial_{x}G_{x}+\partial_{y}\left(\frac{G_{y}}{\gamma}\right)+\partial_{z}\left(\frac{G_{z}}{\gamma}\right) & =-4\pi g\rho',\nonumber \\
\partial_{x}G_{x}+\partial_{y}G_{y}+\partial_{z}G_{z} & =-4\pi g\rho'\gamma,\nonumber \\
\partial_{x}G_{x}+\partial_{y}G_{y}+\partial_{z}G_{z} & =-4\pi g\rho,
\end{align}
where we have used the first of relations (\ref{dGdT}) and the transformation
law for the gravitational mass density (\ref{MassDensity}) as well.
We found therefore that the gravitational field also satisfies the
Gauss law in $S$, namely, 
\begin{equation}
\boldsymbol{\nabla}\cdot\boldsymbol{G}=-4\pi g\rho.\label{DivG}
\end{equation}

Now, let us verify what should be the curl of the gravitational field.
Replacing the coordinate derivatives given in (\ref{Derivs}) in (\ref{RotG'})
and using (\ref{Gprimed}) we get, for each component, 
\begin{align}
\partial_{y}'G_{z}'-\partial_{z}'G_{y}' & =\partial_{y}\left(\frac{G_{z}}{\gamma}\right)-\partial_{y}\left(\frac{G_{y}}{\gamma}\right)=0,\nonumber \\
\partial_{z}'G_{x}'-\partial_{x}'G_{z}' & =\partial_{z}G_{x}-\gamma\left(\partial_{x}+\frac{v}{c^{2}}\partial_{t}\right)\left(\frac{G_{z}}{\gamma}\right)=0,\nonumber \\
\partial_{x}'G_{y}'-\partial_{y}'G_{x}' & =\gamma\left(\partial_{x}+\frac{v}{c^{2}}\partial_{t}\right)\left(\frac{G_{y}}{\gamma}\right)-\partial_{y}G_{x}=0,
\end{align}
that is, 
\begin{equation}
\partial_{y}G_{z}-\partial_{z}G_{y}=0,\qquad\partial_{z}G_{x}-\partial_{x}G_{z}=\frac{v}{c^{2}}\partial_{t}G_{z},\qquad\partial_{x}G_{y}-\partial_{y}G_{x}=-\frac{v}{c^{2}}\partial_{t}G_{y}.\label{RotG}
\end{equation}
Now, remembering that the magnetic field is given in $S$ by (\ref{FieldB}),
we get, 
\begin{equation}
B_{x}=0,\qquad B_{y}=-\frac{v}{c^{2}}G_{z},\qquad B_{z}=\frac{v}{c^{2}}G_{y},\label{Bcomp}
\end{equation}
from which we conclude that 
\begin{equation}
\boldsymbol{\nabla}\times\boldsymbol{G}=-\partial_{t}\boldsymbol{B}.\label{RotG1}
\end{equation}
Notice that the gravitational field is no longer irrotational in $S$,
which express the content of the gravitational analogue of Faraday's
law. Since $\boldsymbol{\nabla}\times\boldsymbol{G}\neq0$, it might
seem that the gravitational field is not conservative anymore. This
is only apparent, however, since we shall see in sections \ref{Sec_GravPot}
and \ref{Sec_EPTensor} that the gravitational fields have, besides
an energy, also a momentum associated, so that the \emph{spacetime
momentum} of the fields is conserved in any inertial reference frame.

Together with (\ref{DivG}), equation (\ref{RotG1}) completes the
set of differential equations for the gravitational field $\boldsymbol{G}$
as measured by $S$. Still remains, however, to find the divergence
and the curl of the magnetic field $\boldsymbol{B}$. This is quite
simple to be found. In fact, since the magnetic field components are
given by (\ref{Bcomp}), we get at once that,
\begin{equation}
\partial_{x}B_{x}+\partial_{y}B_{y}+\partial_{z}B_{z}=\partial_{y}\left(-\frac{v}{c^{2}}G_{z}\right)+\partial_{z}\left(\frac{v}{c^{2}}G_{y}\right)=-\frac{v}{c^{2}}\left(\partial_{y}G_{z}-\partial_{z}G_{y}\right)=0,
\end{equation}
where we made use of the first equation in (\ref{RotG}). Therefore,
we conclude that the divergence of the magnetic field is always zero\footnote{Notice that this represents a strong evidence that classical magnetic
monopoles should not exist, since our argument shows that the vanishing
of the divergence of the magnetic field follows from a direct consequence
of the special relativity theory. Besides, our definition of magnetic
field (which agrees with the usual one) as the field that arises in
order to balance the forces in all inertial reference frames, makes
non-sense when we talk about magnetic monopoles classically, since
every field generated by such a particle would be simply interpreted
as some kind of gravitational (electric) field generated by ordinary
gravitational (electric) charges anyway. }, 
\begin{equation}
\boldsymbol{\nabla}\cdot\boldsymbol{B}=0.
\end{equation}

Finally, let us verify what should be the curl of the magnetic field.
We have, at first, that 
\begin{align}
\partial_{y}B_{z}-\partial_{z}B_{y} & =\partial_{y}\left(\frac{v}{c^{2}}G_{y}\right)-\partial_{z}\left(-\frac{v}{c^{2}}G_{z}\right)=\frac{v}{c^{2}}\left(\partial_{y}G_{y}+\partial_{z}G_{z}\right),\nonumber \\
\partial_{z}B_{x}-\partial_{x}B_{z} & =-\partial_{x}\left(\frac{v}{c^{2}}G_{y}\right)=-\frac{v}{c^{2}}\partial_{x}G_{y},\nonumber \\
\partial_{x}B_{y}-\partial_{y}B_{x} & =\partial_{x}\left(-\frac{v}{c^{2}}G_{z}\right)=-\frac{v}{c^{2}}\partial_{x}G_{z}.
\end{align}
However, from the Gauss law we know that $\partial_{y}G_{y}+\partial_{z}G_{z}=-4\pi g\rho-\partial_{x}G_{x}$
and, hence, replacing the derivatives in $x$ by the derivatives in
$t$ through (\ref{dGdT}), we get the equations, 
\begin{equation}
\partial_{y}B_{z}-\partial_{z}B_{y}=-\frac{4\pi g}{c^{2}}\rho v+\partial_{t}G_{x},\qquad\partial_{z}B_{x}-\partial_{x}B_{z}=-\frac{1}{c^{2}}\partial_{t}G_{y},\qquad\partial_{x}B_{y}-\partial_{y}B_{x}=-\frac{1}{c^{2}}\partial_{t}G_{z},
\end{equation}
which can be written as well in a vector form as, 
\begin{equation}
\boldsymbol{\nabla}\times\boldsymbol{B}=-\frac{4\pi g}{c^{2}}\boldsymbol{j}+\frac{1}{c^{2}}\partial_{t}\boldsymbol{G},
\end{equation}
where $\boldsymbol{j}=\rho\boldsymbol{v}=\rho v\hat{\boldsymbol{x}}$
is the current density.

In conclusion, the differential equations satisfied by the gravitomagnetic
fields in the reference frame $S$ are, 
\begin{align}
\boldsymbol{\nabla}\cdot\boldsymbol{G} & =-4\pi g\rho,\label{MdivG}\\
\boldsymbol{\nabla}\cdot\boldsymbol{B} & =0,\label{MdivB}\\
\boldsymbol{\nabla}\times\boldsymbol{G} & =-\partial_{t}\boldsymbol{B},\label{McurlG}\\
\boldsymbol{\nabla}\times\boldsymbol{B} & =-\frac{4\pi g}{c^{2}}\boldsymbol{j}+\frac{1}{c^{2}}\partial_{t}\boldsymbol{G}.\label{McurlB}
\end{align}
They have the same form of the Maxwell equations of electromagnetic
theory, besides the signs in front of the sources $\rho$ and $\boldsymbol{j}$.
These negative signs, of course, is due to the fact that the gravitational
interaction is always attractive.

In these equations, notice that the only things that matters are the
fields $\boldsymbol{G}$ and $\boldsymbol{B}$ and the densities of
mass and current, $\rho$ and $\boldsymbol{j}$, all quantities being
measured in $S$. However, since the velocity in which $S$ moves
w.r.t. $S'$ is quite arbitrary, we may conclude that the Maxwell
equations should maintain their form in any inertial frame of reference.
This can be in fact proved in another way, by considering the reference
frame $S''$, in the same fashion as was made in section \ref{Sec_TransFields}.
This calculation will be concealed.

This approach shows us that the Maxwell equations have nothing to
do with the nature of the electromagnetic fields $-$ on the contrary,
we showed that they are just a consequence of the covariance enjoyed
by the fields regarding the Lorentz transformations. Therefore, any
other \emph{covariant field} (e.g., the electromagnetic field or any
other else) must satisfy a similar set of equations. This point can
also be evidenced in an interesting work of Heras \cite{Heras}, where
is stated that the Maxwell equations can be deduced uniquely from
the continuity equation\footnote{To be more specific, Heras stated that, if there are two localized
functions $\phi\left(x,y,z,t\right)$ and $\boldsymbol{j}(x,y,z,t)$
that satisfy the continuity equation (\ref{Continuity}), then there
exist two vector functions $\boldsymbol{F}(x,y,z,t)$ and $\boldsymbol{G}(x,y,z,t)$
that satisfy a set of equations similar to that of Maxwell. See \cite{Heras}
for details.}. Since we have not used the continuity equation in our derivation
of the Maxwell equations, and since, on the other hand, the continuity
equation is a direct consequence of the Maxwell equations, this means
that both set of equations are, surprisingly, mathematically equivalent.

\section{Gravitational waves\label{Sec_GravWaves}}

The gravitational Maxwell equations presented in the previous section
consist of a coupled system of linear partial differential equations
for the fields $\boldsymbol{G}$ and $\boldsymbol{B}$. This set of
equations nevertheless can be written in a decoupled form as well.
To see this, we can follow the standard methods employed in the electromagnetic
theory. Thus, let us begin by taking the curl of (\ref{McurlG}):
\begin{align}
\boldsymbol{\nabla}\times\left(\boldsymbol{\nabla}\times\boldsymbol{G}\right) & =\boldsymbol{\nabla}\times\left(-\partial_{t}\boldsymbol{B}\right), & \text{that is,} &  & \boldsymbol{\nabla}\left(\boldsymbol{\nabla}\cdot\boldsymbol{G}\right)-\boldsymbol{\nabla}^{2}\boldsymbol{G} & -\partial_{t}\left(\boldsymbol{\nabla}\times\boldsymbol{B}\right),
\end{align}
where we had used the vector identities (\ref{V3}) and (\ref{V4}).
Using now (\ref{MdivG}) and (\ref{McurlB}), we get, after simplification,
\begin{equation}
\boldsymbol{\nabla}^{2}\boldsymbol{G}-\frac{1}{c^{2}}\partial_{t}^{2}\boldsymbol{G}=-4\pi g\left(\boldsymbol{\nabla}\rho+\frac{1}{c^{2}}\partial_{t}\boldsymbol{j}\right),\label{InhomG}
\end{equation}
which is a decoupled differential equation for the field $\boldsymbol{G}$.
Likewise, taking the curl of (\ref{McurlB}) and using the identities
(\ref{V3}) and (\ref{V4}) again, we get, 
\begin{align}
\boldsymbol{\nabla}\times\left(\boldsymbol{\nabla}\times\boldsymbol{B}\right) & =\boldsymbol{\nabla}\times\left(-\frac{4\pi g}{c^{2}}\boldsymbol{j}+\frac{1}{c^{2}}\partial_{t}\boldsymbol{G}\right) & \text{or,} &  & \boldsymbol{\nabla}\left(\boldsymbol{\nabla}\cdot\boldsymbol{B}\right)-\boldsymbol{\nabla}^{2}\boldsymbol{B} & =-\frac{4\pi g}{c^{2}}\left(\boldsymbol{\nabla}\times\boldsymbol{j}\right)+\frac{1}{c^{2}}\partial_{t}\left(\boldsymbol{\nabla}\times\boldsymbol{G}\right),
\end{align}
and, from the Maxwell equations (\ref{MdivB}) and (\ref{McurlG}),
we obtain, 
\begin{equation}
\boldsymbol{\nabla}^{2}\boldsymbol{B}-\frac{1}{c^{2}}\partial_{t}^{2}\boldsymbol{B}=\frac{4\pi g}{c^{2}}\left(\boldsymbol{\nabla}\times\boldsymbol{j}\right),\label{InhomB}
\end{equation}
which is the decoupled differential equation for the field $\boldsymbol{B}$.

Equations (\ref{InhomG}) and (\ref{InhomB}) are the gravitomagnetic
\emph{inhomogeneous wave equations}. In the absence of matter, that
is, whenever $\rho=0$ and $\boldsymbol{j}=0$, (\ref{InhomG}) and
(\ref{InhomB}) simplify to 
\begin{equation}
\boldsymbol{\nabla}^{2}\boldsymbol{G}-\frac{1}{c^{2}}\partial_{t}^{2}\boldsymbol{G}=0,\qquad\boldsymbol{\nabla}^{2}\boldsymbol{B}-\frac{1}{c^{2}}\partial_{t}^{2}\boldsymbol{B}=0,\label{Wave}
\end{equation}
and we get the gravitomagnetic \emph{homogeneous wave equation}s.

We highlight that (\ref{Wave}) shows us clearly that gravitational
waves propagate with the speed of the light in vacuum. We remark that
this result follows directly from the special theory of relativity
and that this result cannot be derived by making a mere analogy between
gravitation and electromagnetism, since, in this case, there is no
way to guess what should be the gravitational permittivity and permeability
of the vacuum. The recent detection of gravitational waves by the
\noun{ligo} experiments \cite{Abbott2016,Abbott2016b,Abbott2017a,Abbott2017b}
confirmed that gravitational waves indeed propagate with the speed
of light, $c$, which represents a strong argument in favor of our
present theory \textendash{} and that also disproofs any other gravitomagnetic
theory concluding in a different way. 

\section{The gravitomagnetic potentials and their connection with the interaction
energy and momentum\label{Sec_GravPot}}

When a given distribution of matter is at rest, it is very known that
the gravitational field so created is conservative. In fact, this
is ensured by the third Maxwell equation (\ref{McurlG}), which states
that the gravitational field is irrotational whenever there is no
magnetic field and, thus, it is possible to write the gravitational
field $\boldsymbol{G}'$ as the gradient of a\emph{ scalar potential
}function $\phi'$, for instance, as 
\begin{equation}
\boldsymbol{G}'\left(x',y',z'\right)=-\boldsymbol{\nabla}'\phi'\left(x',y',z'\right).
\end{equation}
The physical meaning of the function $\phi'$ is easily found to be
the \emph{interaction energy} by unit mass between the particle and
the gravitational field. In other words, a particle of mass $m$ moving
in a static gravitational field has an interaction energy given by
\begin{equation}
U'\left(x',y',z'\right)=m\phi'\left(x',y',z'\right).\label{U'}
\end{equation}

This picture changes, however, when the source of the gravitational
field is moving. In this case the gravitational field cannot be conservative
in the sense above, since the presence of the magnetic fields prevents
the field from being irrotational, as (\ref{McurlG}) states. Nevertheless,
we shall see in the following that in the frame $S$ there exists
as well a \emph{vector potential} function $\boldsymbol{A}$, which
is associated with the \emph{interaction momentum} between the gravitomagnetic
fields and the particle and that solves the problem. The scalar potential
$\phi$ and the vector potential $\boldsymbol{A}$ form, in fact,
a spacetime vector (the \emph{spacetime potential}) which is proportional
to the spacetime momentum associated with the interaction between
the fields and the particle. This \emph{spacetime interaction momentum}
is a conserved quantity and hence, in this more embracing sense, the
gravitomagnetic field can be regarded as conservative.

We can easily show the existence of the gravitomagnetic potentials
$\phi$ and $\boldsymbol{A}$ directly from the homogeneous Maxwell
equations (\ref{MdivB}) and (\ref{McurlG}) $-$ that is, the Maxwell
equations which do not depend on the sources $\rho$ and $\boldsymbol{j}$.
In fact, the divergenceless property of the magnetic field stated
in the second Maxwell equation (\ref{MdivB}) means, through identity
(\ref{V1}), that there exists a $C^{2}$ class vector function $\boldsymbol{A}$
satisfying the condition 
\begin{equation}
\boldsymbol{B}=\boldsymbol{\nabla}\times\boldsymbol{A}.\label{BcurlA}
\end{equation}

Likewise, replacing (\ref{BcurlA}) into the third Maxwell equation
(\ref{McurlG}) we shall get, after using the previous formula (\ref{BcurlA})
and the identity (\ref{V3}), 
\begin{equation}
\boldsymbol{\nabla}\times\boldsymbol{G}+\partial_{t}\boldsymbol{B}=\boldsymbol{\nabla}\times\boldsymbol{G}+\partial_{t}\left(\boldsymbol{\nabla}\times\boldsymbol{A}\right)=\boldsymbol{\nabla}\times\left(\boldsymbol{G}+\partial_{t}\boldsymbol{A}\right)=0.
\end{equation}
Therefore we can realize, through the identity (\ref{V2}), that there
exists a $C^{2}$ scalar function $\phi$ such that, 
\begin{equation}
\boldsymbol{G}+\partial_{t}\boldsymbol{A}=-\boldsymbol{\nabla}\phi,\qquad\text{that is,}\qquad\boldsymbol{G}=-\boldsymbol{\nabla}\phi-\partial_{t}\boldsymbol{A}.\label{GgradPhi}
\end{equation}

The existence of the gravitomagnetic potentials $\phi$ and $\boldsymbol{A}$
given by (\ref{BcurlA}) and (\ref{GgradPhi}) provides the required
relationship between the gravitomagnetic fields with the potentials. 

The gravitomagnetic potentials, however, are not uniquely determined,
since we can replace $\phi$ and $\boldsymbol{A}$ respectively by
$\phi'=\phi+\partial_{t}f$ and $\boldsymbol{A}'=\boldsymbol{A}+\boldsymbol{\nabla}f$,
where $f$ is any $C^{2}$ scalar function, without any change in
the gravitomagnetic fields. This feature is often called the \emph{gauge
freedom} of the potentials. Hence the covariant theory of gravitation
here outlined has gauge symmetry as the electromagnetic theory. This
feature, which is also a consequence of relativity (and therefore
should be present in any relativistic force field), can be of importance
in the formulation of a quantum version of this theory. 

Let us see now why the potentials $\phi$ and $\boldsymbol{A}$ should
be associated with the interaction energy and momentum between a particle
and the gravitomagnetic fields. To this end, consider first a particle
that is moving in a static gravitational field, say w.r.t. a reference
frame $S'$. The energy of the particle is composed of two terms:
the kinetic energy $K'=\gamma_{u'}mc^{2}$ and the interaction energy
$U'=m\phi'$ as given by (\ref{U'}). Moreover, the particle has a
kinetic momentum given by $\boldsymbol{p}'=\gamma_{u'}m\boldsymbol{u}'.$
Now, take the reference frame $S$ where the distribution of matter
which generated the fields has a velocity $\boldsymbol{v}=v\hat{\boldsymbol{x}}$.
Then we may ask what is the energy and the momentum of the particle
as it moves in the gravitomagnetic fields. The answer is again provided
by the special theory of relativity, since the total energy and momentum
of the particle should be the components of a spacetime vector. This
means that the total energy $E$ and the total momentum $\boldsymbol{P}$
of the particle, as measured by $S$, should be related to its total
energy $E'$ and momentum $\boldsymbol{P}'=\boldsymbol{p}'$, as measured
in $S'$, by the formulæ 
\begin{equation}
E=\gamma E',\qquad P_{x}=\frac{\gamma v}{c^{2}}E',\qquad P_{y}=0,\qquad P_{z}=0.
\end{equation}
However, since the same transformation formulæ holds for the kinetic
energy $K$ and momentum $\boldsymbol{p}$ of the particle, this means
that the same also holds for the interaction energy $U$ and the interaction
momentum $\boldsymbol{V}$. Therefore, the simple fact that the particle
has an interaction energy in $S'$ given by (\ref{U'}) implies that
it has both an interaction energy $U$ and an interaction momentum
$\boldsymbol{V}$ in $S$, which are respectively given by 
\begin{equation}
U\left(x,y,z,t\right)=m\phi\left(x,y,z,t\right),\qquad\boldsymbol{V}\left(x,y,z,t\right)=m\boldsymbol{A}\left(x,y,z,t\right),\label{UV}
\end{equation}
where the potentials $\phi\left(x,y,z,t\right)$ and $\boldsymbol{A}\left(x,y,z,t\right)$
are, 
\begin{equation}
\phi\left(x,y,z,t\right)=\gamma\phi'\left(x',y',z'\right),\qquad\text{and}\qquad\boldsymbol{A}\left(x,y,z,t\right)=\frac{\gamma\boldsymbol{v}}{c^{2}}\phi'\left(x',y',z'\right)=\frac{\boldsymbol{v}}{c^{2}}\phi\left(x,y,z,t\right).\label{Phi}
\end{equation}
(The Lorentz transformations still must be used in order to eliminate
the primed coordinates, of course.) From this we conclude that the
gravitomagnetic potentials are the components of a spacetime vector,
as stated above, since they are proportional to the interaction energy
and momentum. Hence, they must transform, from a reference frame $S$
to another frame $S''$, according to the formulæ 
\begin{equation}
\phi''=\gamma\left(\phi-vA_{x}\right),\qquad A_{x}''=\gamma\left(A_{x}-\frac{v}{c^{2}}\phi\right),\qquad A_{y}''=A_{y},\qquad A_{z}''=A_{z}.
\end{equation}

The results derived above can also be deduced from a variational perspective.
To this end, let us take the relativistic Lagrangian for a particle
moving in a static gravitational field, which is, 
\begin{equation}
\mathcal{L}'=-mc^{2}/\gamma_{u'}-m\phi'.
\end{equation}
In the Lagrangian formulation, remember that the canonical energy
and momentum of the particle are given, respectively, by 
\begin{equation}
\boldsymbol{P}'=\dot{\boldsymbol{\nabla}}'\mathcal{L}'=\gamma_{u'}m\boldsymbol{u}^{\prime},\qquad H'=\dot{\boldsymbol{\nabla}}'\mathcal{L}'\cdot\boldsymbol{u}'-\mathcal{L}'=\gamma_{u'}mc^{2}+m\phi',
\end{equation}
from where the relation between the interaction energy and the scalar
potential becomes plain: $U'=m\phi'$. The action is, of course, given
by 
\begin{equation}
S=\int\mathcal{L}'\mathrm{d}t'=\int\boldsymbol{P}'\cdot\mathrm{d}\boldsymbol{r}'-\int H'\mathrm{d}t'.
\end{equation}

Now we pass to the reference frame $S$, where the gravitational field
is moving, and we ask what happens there. The key point here is to
realize that the action is Lorentz invariant. This means that the
canonical energy and the canonical momentum components must transform
as a spacetime vector. Since the same holds for the kinetic energy
and momentum, we found again that the same must be true for the interaction
energy and momentum. In fact, it is not difficult to show that the
canonical momentum and energy of the particle in the reference frame
$S$, are given respectively by, 
\begin{equation}
\boldsymbol{P}=\gamma_{u}m\boldsymbol{u}+m\boldsymbol{A},\qquad H=\gamma_{u}mc^{2}+m\phi,\label{PH}
\end{equation}
from where (\ref{UV}) and (\ref{Phi}) directly follows. The relationship
between the fields and the potentials can also be recovered from this
variational approach \cite{Landau1980}. Now we need the equation
of motion of the particle in the reference frame $S$, which is provided
by the Euler-Lagrange equations, 
\begin{equation}
\frac{\mathrm{d}}{\mathrm{d}t}\dot{\boldsymbol{\nabla}}\mathcal{L}=\boldsymbol{\nabla}\mathcal{L},\label{EulerLagrange}
\end{equation}
with the transformed Lagrangian, 
\begin{equation}
\mathcal{L}=-mc^{2}/\gamma_{u}-m\phi+m\left(\boldsymbol{A}\cdot\boldsymbol{u}\right).\label{Lag}
\end{equation}
Notice, however, that the force acting of the particle is given by
the time derivative of its kinetic momentum $\boldsymbol{p}$, not
by the time derivative of the canonical momentum $\boldsymbol{P}$.
Thus, the Euler-Lagrange equations provide the force only indirectly.
Nonetheless, it is easy to verify from (\ref{PH}) that 
\begin{equation}
\frac{\mathrm{d}\boldsymbol{p}}{\mathrm{d}t}=\frac{\mathrm{d}\boldsymbol{P}}{\mathrm{d}t}-m\frac{\mathrm{d}\boldsymbol{A}}{\mathrm{d}t}.\label{Kforce}
\end{equation}
Therefore, form the Lagrangian (\ref{Lag}), we get that 
\begin{equation}
\frac{\mathrm{d}\boldsymbol{P}}{\mathrm{d}t}=\boldsymbol{\nabla}\mathcal{L}=-m\boldsymbol{\nabla}\phi+m\boldsymbol{\nabla}\left(\boldsymbol{A}\cdot\boldsymbol{u}\right),
\end{equation}
and remembering that the derivatives are carried out for constant
$\boldsymbol{u}$ we get, after using the vector identity (\ref{V8}),
\begin{equation}
\boldsymbol{\nabla}\mathcal{L}=-m\boldsymbol{\nabla}\phi+m\left(\boldsymbol{u}\cdot\boldsymbol{\nabla}\right)\boldsymbol{A}+m\boldsymbol{u}\times\left(\boldsymbol{\nabla}\times\boldsymbol{A}\right).\label{GradL}
\end{equation}
On the other hand, the derivative $\mathrm{d}\boldsymbol{A}/\mathrm{d}t$
is, 
\begin{equation}
\frac{\mathrm{d}\boldsymbol{A}}{\mathrm{d}t}=\partial_{t}\boldsymbol{A}+\left(\boldsymbol{u}\cdot\boldsymbol{\nabla}\right)\boldsymbol{A},\label{dAdt}
\end{equation}
and, hence, inserting (\ref{GradL}) and (\ref{dAdt}) on (\ref{Kforce}),
we find that the force acting on the particle is, 
\begin{equation}
\boldsymbol{F}=m\left[-\left(\boldsymbol{\nabla}\phi+\partial_{t}\boldsymbol{A}\right)+\boldsymbol{u}\times\left(\boldsymbol{\nabla}\times\boldsymbol{A}\right)\right].
\end{equation}

Now, since, by definition, the gravitational force is the part of
the gravitomagnetic force which does not depend on the particle velocity,
while the magnetic force is that part which does depend, we conclude
that the total force acting on the particle in the reference frame
$S$ can be written in the Lorentz form as 
\begin{equation}
\boldsymbol{F}=m\left(\boldsymbol{G}+\boldsymbol{u}\times\boldsymbol{B}\right),
\end{equation}
where the gravitational and magnetic fields, $\boldsymbol{G}$ and
$\boldsymbol{B}$, are related to the scalar and vector potentials,
$\phi$ and $\boldsymbol{A}$, respectively by 
\begin{equation}
\boldsymbol{G}=-\boldsymbol{\nabla}\phi-\partial_{t}\boldsymbol{A},\qquad\boldsymbol{B}=\boldsymbol{\nabla}\times\boldsymbol{A}.
\end{equation}
This variational approach provides another deduction of (\ref{BcurlA})
and (\ref{GgradPhi}) and of the results obtained in section \ref{Sec_Force}.

Before closing this section, we would like to make some remarks about
the association of the vector potential $\boldsymbol{A}$ with the
interaction momentum. In the case of the electromagnetic theory, this
association is by no means new and the first connections remount back
to the original works of Maxwell \cite{Maxwell1865}, although this
connection was getting lost in the course of time \cite{Semon1996,Griffiths2012}.
Even today, this subject is controversial and many physicists believe
that no physical meaning can be associated with the vector potential
$-$ their criticisms are commonly based on the fact that the vector
potential has a gauge freedom, while it is hard to explain this feature
in terms of the momentum. This argument, however, is flawed. There
is no doubt that the kinetic momentum of a particle does not allow
a gauge freedom, since it is only a function of its velocity, however,
the same could be said about the kinetic energy. For the interaction
energy, on the other hand, we can set the ground state as we please
and since the interaction energy is proportional to the scalar potential,
we have a gauge freedom as well. Therefore, the same arguments can
be applied to interaction momentum, and there is no reason to prevent
the vector potential of being associated with the interaction momentum.
Moreover, the theory of relativity sets momentum and energy on an
equal foot, whence this association is mandatory instead of optional.
Finally, we should remember that in quantum mechanics this association
is important to explain, for instance, the Aharonov-Bohm effect \cite{Aharonov1959,GriffithsQM}.

\section{The energy and the momentum stored in the gravitomagnetic fields\label{Sec_EPTensor}}

We have seen in the previous section that a particle moving in a non-static
gravitational field has both an interaction energy as well as an interaction
momentum $-$ these quantities are directly related to the gravitomagnetic
potentials $\phi$ and $A$ through (\ref{UV}) and (\ref{Phi}).
In this section, we shall show that the fields themselves store energy
and momentum. The energy and momentum contained in the gravitomagnetic
fields can be regarded as distributed throughout the space, thus,
we can talk about the energy and momentum densities as well as the
energy and momentum fluxes. In the one hand, we shall see that, if
there are no mass or current in a given region of space, then the
energy and the momentum of the fields will be conserved there, so
that they will satisfy a kind of continuity equation. On the other
hand, if matter is present in the space, then we shall see that the
fields must transmit energy and momentum to the matter, so that the
continuity equation will be replaced by an appropriated balance equation.

Let us begin our analysis by deducing what should be the energy stored
in the gravitomagnetic fields. This can be made in the following mathematical
way: first we dot the third Maxwell equation (\ref{McurlG}) with
$\boldsymbol{G}$ and then the fourth Maxwell equation (\ref{McurlB})
with $\boldsymbol{B}$: 
\begin{equation}
\boldsymbol{B}\cdot\left(\boldsymbol{\nabla}\times\boldsymbol{G}\right)=-\boldsymbol{B}\cdot\partial_{t}\boldsymbol{B},\qquad\boldsymbol{G}\cdot\left(\boldsymbol{\nabla}\times\boldsymbol{B}\right)=-\frac{4\pi g}{c^{2}}\boldsymbol{G}\cdot\boldsymbol{j}+\frac{1}{c^{2}}\boldsymbol{G}\cdot\partial_{t}\boldsymbol{G}.
\end{equation}
Taking the difference of these two equations and using the vector
identity (\ref{V5}), we obtain, 
\begin{equation}
\boldsymbol{\nabla}\cdot\left(\boldsymbol{G}\times\boldsymbol{B}\right)=\frac{4\pi g}{c^{2}}\boldsymbol{G}\cdot\boldsymbol{j}-\frac{1}{c^{2}}\boldsymbol{G}\cdot\partial_{t}\boldsymbol{G}-\boldsymbol{B}\cdot\partial_{t}\boldsymbol{B},
\end{equation}
and, dividing it by $4\pi g/c^{2}$, using the identity (\ref{V6}),
and rearranging, we get, 
\begin{equation}
\boldsymbol{\nabla}\cdot\left[\frac{c^{2}}{4\pi g}\left(\boldsymbol{G}\times\boldsymbol{B}\right)\right]+\partial_{t}\left[\frac{1}{8\pi g}\left(\boldsymbol{G}\cdot\boldsymbol{G}+c^{2}\boldsymbol{B}\cdot\boldsymbol{B}\right)\right]=\boldsymbol{G}\cdot\boldsymbol{j}.\label{RateG}
\end{equation}
This equation represents the time rate per unit volume in which the
gravitomagnetic fields deliver energy to the matter. In fact, by introducing
the quantities 
\begin{equation}
\boldsymbol{S}=-\frac{c^{2}}{4\pi g}\left(\boldsymbol{G}\times\boldsymbol{B}\right),\qquad\text{and}\qquad\mathcal{E}=-\frac{1}{8\pi g}\left(\boldsymbol{G}\cdot\boldsymbol{G}+c^{2}\boldsymbol{B}\cdot\boldsymbol{B}\right),\label{Poynting}
\end{equation}
we may realize that (\ref{RateG}) can be rewritten as 
\begin{equation}
\boldsymbol{\nabla}\cdot\boldsymbol{S}+\partial_{t}\mathcal{E}=-\boldsymbol{G}\cdot\boldsymbol{j}.\label{RateG1}
\end{equation}
The quantity $-\boldsymbol{G}\cdot\boldsymbol{j}$, however, is nothing
but the power (per unit volume) exerted by the fields on the matter
$-$ the minus sign means that the fields lost energy in this process.
Consequently, the quantity $\mathcal{E}$ represents the energy density
of the gravitational field and the vector $\boldsymbol{S}$ consists
in the flux of this energy (this vector is the gravitational analogue
of the Poynting vector found in the electromagnetic theory). Equation
(\ref{RateG}) \textendash{} or its equivalent, equation (\ref{RateG1})
\textendash{} states that in the presence of matter the energy of
the gravitational fields, alone, is not conserved anymore. Only the
total energy, that is, the energy of the fields plus the energy of
the matter, is conserved.

Furthermore, we should remark that the energy of the gravitational
fields is always negative \textendash{} contrasting with the energy
of the electromagnetic fields, which is always positive. This feature
is sometimes subject to some criticism, since it is not clear at the
first sight what a field with negative energy means. However, we argue
here that this feature is not a problem at all and, on the contrary,
that the negativeness of the gravitational energy is ultimately a
consequence of the gravity being always attractive. To see why, take
the simple example of a test particle initially at rest on the gravitational
field of a spherical body. As the particle begins to move due to the
gravitational interaction, it draws energy from the gravitational
field, which is converted into kinetic energy (an always positive
quantity). Hence, since the total energy of the particle should be
conserved, its potential energy must decrease in this process. But
for the potential energy to decrease while there is an attraction
between the two bodies, it is necessary that this potential energy
be negative, if we agree to set the zero of energy at infinity. Therefore,
the negativeness of the gravitational field is only a manifestation
of the attractive character of gravity. This feature seems to be also
important to explain why a quantum gravity theory can be formulated,
in the approximation of weak fields, by the exchange of spin-1 particles
\textendash{} see section \ref{Sec-Conclusions} for more details.

Let us now concern ourselves with the momentum stored in the gravitomagnetic
fields. Here we can proceed by crossing (from the left) the third
Maxwell equation (\ref{McurlG}) with $\boldsymbol{G}$ and then crossing
(from the right) the fourth Maxwell equation (\ref{McurlB}) with
$\boldsymbol{B}$ in order to get, 
\begin{equation}
\boldsymbol{G}\times\left(\boldsymbol{\nabla}\times\boldsymbol{G}\right)=-\boldsymbol{G}\times\partial_{t}\boldsymbol{B},\qquad\left(\boldsymbol{\nabla}\times\boldsymbol{B}\right)\times\boldsymbol{B}=-\frac{4\pi g}{c^{2}}\left(\boldsymbol{j}\times\boldsymbol{B}\right)+\frac{1}{c^{2}}\left(\partial_{t}\boldsymbol{G}\times\boldsymbol{B}\right).
\end{equation}
Thus, multiplying the second equation above by $-c^{2}$ and summing
with the first, we get, after we use the anti-commuting property of
the vector product and the identity (\ref{V6}),
\begin{equation}
\boldsymbol{G}\times\left(\boldsymbol{\nabla}\times\boldsymbol{G}\right)+c^{2}\left[\boldsymbol{B}\times\left(\boldsymbol{\nabla}\times\boldsymbol{B}\right)\right]=4\pi g\left(\boldsymbol{j}\times\boldsymbol{B}\right)-\partial_{t}\left(\boldsymbol{G}\times\boldsymbol{B}\right).
\end{equation}
Besides, from the vector identity (\ref{V7}) we can rewrite this
as, 
\begin{equation}
\partial_{t}\left(\boldsymbol{G}\times\boldsymbol{B}\right)+\frac{1}{2}\boldsymbol{\nabla}\left[\boldsymbol{G}\cdot\boldsymbol{G}+c^{2}\boldsymbol{B}\cdot\boldsymbol{B}\right]-\left[\left(\boldsymbol{G}\cdot\boldsymbol{\nabla}\right)\boldsymbol{G}+c^{2}\left(\boldsymbol{B}\cdot\boldsymbol{\nabla}\right)\boldsymbol{B}\right]=4\pi g\left(\boldsymbol{j}\times\boldsymbol{B}\right),
\end{equation}
and, finally, from the tensor identity (\ref{VT}), we get,
\begin{equation}
\partial_{t}\left(\boldsymbol{G}\times\boldsymbol{B}\right)+\frac{1}{2}\boldsymbol{\nabla}\left[\boldsymbol{G}\cdot\boldsymbol{G}+c^{2}\boldsymbol{B}\cdot\boldsymbol{B}\right]-\left[\boldsymbol{\nabla}\cdot\left(\boldsymbol{G}\otimes\boldsymbol{G}\right)+c^{2}\boldsymbol{\nabla}\cdot\left(\boldsymbol{B}\otimes\boldsymbol{B}\right)\right]=4\pi g\left(\rho\boldsymbol{G}+\boldsymbol{j}\times\boldsymbol{B}\right),\label{RateB}
\end{equation}
where we used the first and second Maxwell equations (\ref{MdivG})
and (\ref{MdivB}) as well. Equation (\ref{RateB}) can also be written
in a compact form as 
\begin{equation}
\partial_{t}\left(\boldsymbol{S}/c^{2}\right)+\boldsymbol{\nabla}\cdot\mathsf{M}=-\left(\rho\boldsymbol{G}+\boldsymbol{j}\times\boldsymbol{B}\right),\label{RateB1}
\end{equation}
by invoking the vector $\boldsymbol{S}$, defined by (\ref{Poynting})
and introducing the \emph{gravitational Maxwell stress-tensor}, 
\begin{equation}
\mathsf{M}=-\frac{1}{8\pi g}\left[\left(\boldsymbol{G}\cdot\boldsymbol{G}+c^{2}\boldsymbol{B}\cdot\boldsymbol{B}\right)\mathsf{I}-2\left(\boldsymbol{G}\otimes\boldsymbol{G}+c^{2}\boldsymbol{B}\otimes\boldsymbol{B}\right)\right],\label{MTensor}
\end{equation}
where $\mathsf{I}$ is the identity tensor. Equation (\ref{RateB})
\textendash{} or its equivalent, equation (\ref{RateB1}) \textendash{}
represents the time rate per unit volume in which the fields delivers
momentum to the matter. In fact, the quantity $-\left(\rho\boldsymbol{G}+\boldsymbol{j}\times\boldsymbol{B}\right)$
is nothing but the opposite of the Lorentz force per unit volume,
$\boldsymbol{F}=\rho\boldsymbol{G}+\boldsymbol{j}\times\boldsymbol{B}$.
Therefore, in the presence of matter, the momentum of the fields is
no longer conserved $-$ only the total momentum, that is, the momentum
of the fields plus the momentum of the matter, is conserved.

Notice further that the vector $\boldsymbol{S}$ plays a dual role
in those formulæ: in (\ref{RateG1}) it represents the flux of the
gravitational energy, while $\boldsymbol{S}/c^{2}$ in the formula
(\ref{RateB1}) represents the momentum density of the fields. This
symmetry is not a matter of coincidence: is a consequence of special
relativity that the flux of energy and the momentum density are related
in this way.

\section{A manifestly covariant approach\label{Sec_Covariant}}

In this section we shall present another derivation of the previous
results, but following now a spacetime perspective. This means that
all the results and equations will be written in a tensor form, that
is, in a manifestly covariant fashion.

Let us begin by showing the existence of a magnetic gravitational
force directly from this spacetime approach. The key point here is
to replace the concept of ordinary force $\boldsymbol{F}$ by that
one of spacetime force $\varGamma^{\mu}$. Before doing so we should
highlight, however, that in a pure spacetime approach there is no
clear separation between gravitational and magnetic forces. In fact,
this separation is ultimately a matter of convention $-$ it was introduced
first in the electromagnetic theory mainly due to historical reasons,
namely, due to the chronological order on which the electric and magnetic
phenomena were discovered. By this reason, it will be necessary at
the end of our calculations to return to the concept of ordinary force
in order to make the separation of the gravitomagnetic forces into
gravitational and magnetic ones.

In the spacetime framework of special relativity, the spacetime force
$\varGamma^{\mu}$ is defined as the derivative of the spacetime momentum
$p^{\mu}$ w.r.t. the proper time $\tau$,

\begin{equation}
\varGamma^{\mu}=\mathrm{d}p^{\mu}/\mathrm{d}\tau,
\end{equation}
as well as in terms of the spacetime acceleration \cite{FeynmanLectures2013,GriffithsE,Purcell,Landau1980,Pauli1981},
\begin{equation}
\varGamma^{\mu}=ma^{\mu}.
\end{equation}
Once fixed a reference frame $S$, the spacetime force components
are related to the power and ordinary force through the relations,
\begin{equation}
\varGamma^{0}=\gamma_{u}W/c,\qquad\varGamma^{1}=\gamma_{u}F_{x},\qquad\varGamma^{2}=\gamma_{u}F_{y},\qquad\varGamma^{3}=\gamma_{u}F_{z}.\label{4F}
\end{equation}

Now, let us consider a static gravitational force acting on a given
particle, as measured in the proper frame $S'$ of the gravitational
source. Then the components of the spacetime force are given, in this
reference frame, by 
\begin{equation}
\varGamma^{\prime0}=\gamma_{u'}W'/c,\qquad\varGamma^{\prime1}=\gamma_{u'}F{}_{x}^{\prime},\qquad\varGamma^{\prime2}=\gamma_{u'}F{}_{y}^{\prime},\qquad\varGamma^{\prime3}=\gamma_{u'}F{}_{z}^{\prime},
\end{equation}
where $W'$, $F_{x}^{\prime}$, $F_{y}^{\prime}$ and $F_{z}^{\prime}$
do not depend on the particle velocity $\boldsymbol{u}'$, since the
force is purely gravitational in $S'$.

The spacetime force, in contrast to the usual force, is a spacetime
vector, whence their components transform, from $S'$ to $S$, according
to, 
\begin{equation}
\varGamma^{0}=\gamma_{v}\left(\varGamma{}^{\prime0}+\frac{v}{c}\varGamma^{\prime1}\right),\quad\varGamma^{1}=\gamma_{v}\left(\varGamma{}^{\prime1}+\frac{v}{c}\varGamma^{\prime0}\right),\quad\varGamma^{2}=\varGamma{}^{\prime2},\quad\varGamma^{3}=\varGamma{}^{\prime3}.
\end{equation}
From this we can verify that the power and the usual force components,
as measured by $S$, are related to the power and the respective components
of the force in the reference frame $S'$ by, 
\begin{equation}
W=\frac{\gamma_{v}\gamma_{u'}}{\gamma_{u}}\left(W'+vF_{x}^{\prime}\right),\qquad F_{x}=\frac{\gamma_{v}\gamma_{u'}}{\gamma_{u}}\left(F_{x}^{\prime}+\frac{vW'}{c^{2}}\right),\qquad F_{y}=\frac{\gamma_{u'}}{\gamma_{u}}F_{y}^{\prime},\qquad F_{z}=\frac{\gamma_{u'}}{\gamma_{u}}F_{z}^{\prime}.
\end{equation}
However, it can be deduced from (\ref{LTV}) that the $\gamma$-factors
are related by the formula, 
\begin{equation}
\frac{\gamma_{u'}}{\gamma_{u}}=\gamma_{v}\left(1-\frac{u_{x}v}{c^{2}}\right),
\end{equation}
from which we get, after simplification, the expressions, 
\begin{align}
W & =\gamma_{v}^{2}\left(1-\frac{u_{x}v}{c^{2}}\right)\left(W'+vF_{x}^{\prime}\right),\nonumber \\
F_{x} & =\gamma_{v}^{2}\left(1-\frac{u_{x}v}{c^{2}}\right)\left(F_{x}^{\prime}+\frac{vW'}{c^{2}}\right)=F_{x}'+\frac{\gamma_{v}v}{c^{2}}\left(F_{y}'u_{y}+F_{z}'u_{z}\right),\nonumber \\
F_{y} & =\gamma_{v}\left(1-\frac{u_{x}v}{c^{2}}\right)F_{y}^{\prime},\nonumber \\
F_{z} & =\gamma_{v}\left(1-\frac{u_{x}v}{c^{2}}\right)F_{z}^{\prime},\label{LTF2}
\end{align}
 where we used in the second equation the identity $W'=\boldsymbol{F}'\cdot\boldsymbol{u}'$
and also the transformation formulæ for the velocity components. Thus
we recovered the force transformations (\ref{LTF1}) presented in
section \ref{Sec_Force}, which enable us finally to separate the
gravitomagnetic force into the gravitational and magnetic parts, leading
us again to the formulæ (\ref{FG}) and (\ref{FM}).

All other results can be expressed in a manifestly covariant way as
well. The mass and current densities, in this spacetime description,
compose into a \emph{spacetime current density}, 
\begin{equation}
J^{\mu}=\rho_{0}w^{\mu},
\end{equation}
where $\rho_{0}$ is the proper mass density and $w^{\mu}=\mathrm{d}x^{\mu}/\mathrm{d}\tau$
is the spacetime velocity. The time component of this spacetime vector
is proportional to the mass density: $J^{0}=\rho_{0}\gamma c=\rho c$,
while its space components are related to the current density: $J^{a}=\rho_{0}\gamma u^{a}=\rho u^{a}=j^{a}$.
We can also verify that the continuity equation (\ref{Continuity})
is written just as a spacetime divergence, 
\begin{equation}
\partial_{\mu}J^{\mu}=0.
\end{equation}

In section \ref{Sec_GravPot}, we have seen that a particle moving
in a gravitomagnetic field has both an interaction energy as an interaction
momentum. These two quantities can also be unified into a spacetime
vector, namely, the \emph{spacetime interaction momentum}, 
\begin{equation}
V^{\mu}=mA^{\mu}.
\end{equation}
This means that the scalar potential $\phi$ and the vector potential
$\boldsymbol{A}$ also form a spacetime vector: the \emph{spacetime
potential} $A^{\mu}$, whose components are 
\begin{equation}
A^{0}=\phi/c,\qquad A^{1}=A_{x},\qquad A^{2}=A_{y},\qquad A^{3}=A_{z}.
\end{equation}
From this we can plainly see that the vector potential is associated
with the interaction momentum as the scalar potential is associated
with the interaction energy.

Moreover, it follows that the gravitomagnetic fields can be found
through the derivatives of the spacetime potential $A^{\mu}$ w.r.t.
the spacetime coordinates, 
\begin{equation}
G^{\mu\nu}=\partial^{\mu}A^{\nu}-\partial^{\nu}A^{\mu}.
\end{equation}
Therefore, the gravitomagnetic fields components form a two-rank anti-symmetric
spacetime tensor, the \emph{spacetime gravitomagnetic field} $G^{\mu\nu}$.
In a matrix representation this tensor can be written as 
\begin{equation}
G^{\mu\nu}=\left(\begin{array}{cccc}
0 & -G_{x}/c & -G_{y}/c & -G_{z}/c\\
G_{x}/c & 0 & -B_{z} & B_{y}\\
G_{y}/c & B_{z} & 0 & -B_{x}\\
G_{z}/c & -B_{y} & B_{x} & 0
\end{array}\right).
\end{equation}

The \emph{spacetime Lorentz force} can also be expressed in terms
of the gravitomagnetic tensor as follows:
\begin{equation}
\varGamma^{\mu}=mw_{\nu}G^{\mu\nu}.
\end{equation}
Notice that the time component of the spacetime Lorentz force is proportional
to the power delivered by the gravitomagnetic fields to the matter,
while its space components are proportional to the usual Lorentz force.
Notice further that, since $\varGamma^{\mu}=ma^{\mu}$, the motion
of a particle in a gravitomagnetic field does not depend on its mass
$-$ this property is the content of Einstein's equivalence principle
and it is one of the most important features of the gravitomagnetic
interaction.

Continuing with the manifestly covariant description, we can also
verify that the four gravitational Maxwell equations become given
by just two tensor equations: the inhomogeneous Maxwell equations
(\ref{MdivG}) and (\ref{McurlB}) can be written as 
\begin{equation}
\partial_{\mu}G^{\mu\nu}=-\frac{4\pi}{c^{2}}J^{\nu},
\end{equation}
while the homogeneous Maxwell equations (\ref{MdivB}) and (\ref{McurlG}),
become given by 
\begin{equation}
\partial_{\alpha}G_{\beta\gamma}+\partial_{\beta}G_{\gamma\alpha}+\partial_{\gamma}G_{\alpha\beta}=0.
\end{equation}

The balance equations for the energy and momentum stored in the fields,
expressed by (\ref{RateG1}) and (\ref{RateB1}), can also be written
as in a simple tensor form as 
\begin{equation}
\partial_{\nu}T^{\mu\nu}=-J_{\nu}G^{\mu\nu}=-f^{\mu},\label{CovariantTransfer}
\end{equation}
where $f^{\mu}=\rho w_{\nu}G^{\mu\nu}$ is the density of the spacetime
Lorentz force acting on the matter and 
\begin{equation}
T^{\mu\nu}=\frac{c^{2}}{4\pi g}\left(\eta_{\alpha\beta}G^{\mu\alpha}G^{\nu\beta}-\frac{1}{4}\eta^{\mu\nu}G_{\alpha\beta}G^{\alpha\beta}\right),
\end{equation}
is the \emph{spacetime momentum-flux }of the field, so that (\ref{CovariantTransfer})
represents the transfer of the spacetime momentum from the fields
to the matter. When expressed in a matrix form, $T^{\mu\nu}$ becomes,
\begin{equation}
T^{\mu\nu}=\left(\begin{array}{cccc}
\mathcal{E} & S^{1}/c & S^{2}/c & S^{3}/c\\
S^{1}/c & M^{11} & M^{12} & M^{13}\\
S^{2}/c & M^{21} & M^{22} & M^{23}\\
S^{3}/c & M^{31} & M^{32} & M^{33}
\end{array}\right),
\end{equation}
from which we plainly see that $T^{\mu\nu}$ is a two-rank symmetric
tensor. Notice further that we can introduce a spacetime momentum-flux
associated with the matter through 
\begin{equation}
t^{\mu\nu}=\rho w^{\mu}w^{\nu}.
\end{equation}
From this we see that the spacetime force acting on the matter can
be obtained through derivatives:
\begin{equation}
\partial_{\nu}t^{\mu\nu}=\partial_{\nu}\left(\rho w^{\mu}w^{\nu}\right)=w^{\nu}\partial_{\nu}\left(\rho w^{\mu}\right)=\frac{\mathrm{d}\pi^{\mu}}{\mathrm{d}\tau}=f^{\mu},
\end{equation}
where $\pi^{\mu}$ are the components of the spacetime momentum density
and we used the identity $\mathrm{d}/\mathrm{d}\tau=w^{\mu}\partial_{\mu}$.
Substituting this into (\ref{CovariantTransfer}), we get promptly
that, 
\begin{equation}
\partial_{\mu}\left(T^{\mu\nu}+t^{\mu\nu}\right)=0,
\end{equation}
which plainly shows the conservation law for the total energy and
momentum of the system $-$ fields plus matter.

Finally, we would like to remark that the analysis above allow us
to make a point about the meaning of the spacetime momentum-flux tensor
$\mathcal{T}^{\mu\nu}$ that is present in the Einstein field equations,
\begin{equation}
\mathcal{G}^{\mu\nu}=\kappa\mathcal{T}^{\mu\nu}.
\end{equation}
This quantity $\mathcal{T}^{\mu\nu}$ should not be regarded as the
spacetime momentum-flux tensor of the\emph{ gravitational field }itself
\textendash{} as one might think at first sight \textendash{} but,
on the contrary, $\mathcal{T}^{\mu\nu}$ should be associated with
the spacetime momentum-flux of the \emph{matter} that generates the
field. The reason is that in general relativity gravity is not regarded
as a force anymore, but it is only a consequence of the curvature
of the spacetime. Therefore, there can be no real transfer of momentum
and energy from the gravitational fields to the matter and the momentum-flux
tensor of the gravitomagnetic field alone must be trivially null (this
is also in accordance with the fact that in general relativity $T^{\mu\nu}$
is considered null in the regions where there is no matter). This
interpretation means that, in the last instance, the source of the
gravitational fields \textendash{} in both the special as in the general
theory of relativity \textendash{} should be regarded as the matter
and that the gravitational magnetic effects are due to the motion
of the matter\footnote{This also suggests that gravitational waves should not generate themselves
gravitational fields, since they are only massless perturbations of
the spacetime itself. The energy of gravitational waves only acquires
a meaning when they are interpreted in terms of gravitomagnetic fields
propagating in a background flat spacetime.} \textendash{} see also the footnote \ref{FootnoteT}. Now, we can
raise the following question: Why is there a non-null spacetime momentum-flux
tensor in the present theory (which is an approximation of general
relativity) if it is identically null in general relativity? The answer
to this question is a conceptual matter. In general relativity the
gravitational momentum-flux $T^{\mu\nu}$ is null but the spacetime
is curved; when we pass to a description where the spacetime is regarded
as flat, the effects of the curvature of the spacetime become described
by actions of force-fields and, in the same way, the gravitational
field acquire a proper energy and momentum in this interpretation.
In other words, we might say that, for matter curve the spacetime
some energy should be spent: all the energy stored in the gravitational
field (according to a flat spacetime perspective), is spent on the
bending of the spacetime, so that we end with a curved spacetime description
where, now, gravity is no longer a force-field and there is no gravitational
proper energy associated with it anymore.

\section{Differences and similarities between gravity and electricity\label{Sec-Diferences}}

Finally, we shall comment some differences between gravity and electricity
which evince their different nature. We have seen that the special
theory of relativity requires that gravity and electricity should
share many properties in common $-$ thus, the existence of magnetic
fields, the conservation of mass (electric charge), the continuity
equation, the Maxwell equations, the existence of gravitomagnetic
(electromagnetic) waves that propagates in the empty space with light's
speed, among others, are all results that depend uniquely on the covariance
of these fields regarding the Lorentz transformations. Other similarities,
between these interactions, however, are not derived from covariance
requirements, for instance, that both forces satisfy an inverse square
law and that they also depend on the product of the interacting masses
(charges).

Gravity and electricity have three main differences, although. The
first and most obvious difference is that the electric force is incredibly
greater than the gravitational one. The second difference, and indeed
the most dramatic one, is the connection that exists between gravity
and geometry, as Einstein's theory states $-$ it seems that no similar
relationship exists regarding the electromagnetic theory. The third
difference and perhaps the most inexplicable one is that the gravitational
interaction is always attractive, while the electric forces can be
either attractive or repulsive. In other words, while the electric
charges can be either positive or negative, the gravitational charge
\textendash{} that is, the gravitational mass \textendash{} is found
to be only of one type. The existence of only one kind of gravitational
mass that always attract themselves implies huge differences between
gravity and electricity. In the following we shall cite some of them.

First of all, the existence of an always attractive gravitational
force is the true reason why the gravitational force overcomes the
electric force at large distances: while the electric fields are all
neutralized by the existence of positive and negative charges, the
gravitational force only enhances itself. Besides, this also prevents
us of talking about gravitational polarization charges, since there
is no way to form such things with one kind of charge only (hence,
there can be no distinction between the gravitational Maxwell equations
in the matter and in the vacuum). Another difference can be seen if
we consider some gravitational analogues of electric capacitors $-$
for instance, a binary system consisting of a black hole plus a swallowing
star. The fact that the gravitational interaction is always attractive
implies that this gravitational system behave differently than its
electromagnetic analogue. In fact, while in a discharging capacitor
the electrons of the plate move in order to decrease the potential,
so that the system reaches at some point to the equilibrium state,
in the gravitational case the black hole (capacitor) continuously
sucks the material of the star (plate), so that its mass increases
while the mass of the star is depreciated and, hence, no state of
equilibrium can be reached $-$ the star is doomed to be completely
swallowed by the black hole.

Another difference arises from the irradiating fields. In fact, while
an electromagnetic dipole can contribute to the irradiated fields,
the same cannot happen with the gravitomagnetic dipoles. The reason
is that, in the first case, the gravitational dipole moment,
\begin{equation}
\boldsymbol{d}=\int\rho(\boldsymbol{r})\boldsymbol{r}\mathrm{d}V,
\end{equation}
always vanishes in the center-of-mass frame and, in the second case,
the gravitational magnetic dipole,
\begin{equation}
\boldsymbol{m}=\int\rho(\boldsymbol{r})\boldsymbol{u}\times\boldsymbol{r}\mathrm{d}V,
\end{equation}
is just the angular momentum of the system w.r.t. the center of mass
frame and hence it is also conserved. The conclusion is that gravitational
dipoles cannot irradiate: the first contribution should come from
the quadrupoles terms onward (notice that the same property is valid
in Einstein's theory of gravitation \cite{GriffithsE,Purcell,Wald1984,Landau1980,Thorne1980,Thorne1987}).

Finally, we want to highlight that although the gravitational interaction
between two particles is always attractive, there is still an open
question regarding the gravitational interaction between a particle
and an antiparticle (some arguments defending that particles and antiparticles
should repel each other are presented in \cite{Costella1997,Villata2011}).
We remark that although the concept of antiparticle is usually introduced
in the scope of quantum mechanics, their existence can be evidenced
as well from classical point of view, for instance, through the so-called
\emph{Extended Theory of Relativity} \cite{Bilaniuk1962,Antippa1973,Recami1986,Sutherland1986,Vieira2012,Hill2012}
$-$ an extension of the special relativity theory that describes
antiparticles and tachyons\footnote{Proposals to extend the special theory of relativity to describe tachyons
and antiparticles were presented by several authors since the sixties,
especially by Bilaniuk, Deshpande and Sudarshan \cite{Bilaniuk1962},
Antippa and Everett \cite{Antippa1973}, Recami \cite{Recami1986},
Sutherland and Shepanski \cite{Sutherland1986}, among others. In
recent years, some interest in this field was renewed after the works
of Vieira \cite{Vieira2012} and Hill and Cox \cite{Hill2012}, which
independently proposed the same extended Lorentz transformations.
It seems now that there is some consensus about the correctness of
these extended Lorentz transformations, at least in two dimensions
\cite{Grushka2013,Peacock2014,Hill2014,Jin2015}. A consistent extension
of the theory of relativity in higher dimensions is, however, still
an open problem.}. Let us present briefly the lines of reasoning that support the possibility
of a gravitational repulsion between matter and antimatter. Remember
that in the \emph{Stueckelberg-Feynman} interpretation \cite{Feynman1949,Stueckelberg1941},
antiparticles are thought as particles that travel back in time. In
fact, in a spacetime description the energy of a retrograde particle
should be negative, since in this case we have the relation, 
\begin{equation}
E=-\sqrt{m^{2}c^{4}+p^{2}c^{2}}.
\end{equation}
A particle with negative energy that travels back in time should be,
although, actually observed as an ordinary particle with positive
energy, since any observer measures the time from the past to the
future In this process of measurement, however, some properties of
the particle must be reversed, in particular its electric charge \textendash{}
this is in fact the content of the Stueckelberg-Feynman principle.
The reversion of the electric charge can be proved through the \textsc{cpt}
theorem or in terms of the extended Lorentz transformations (from
which the \textsc{cpt} theorem can be classically proved \cite{Recami1986,Vieira2012}).
Now, the same argument that implies the reversal of the electric charge
also must apply to the gravitational mass, since we had shown that
both concepts are quite analogues. Hence, this would mean that the
``gravitational charge'' of antiparticles are negative, which suggests
a repulsive force between a particle and an antiparticle. This conclusion,
if confirmed, could be of importance in cosmology, since it might
explain the observed asymmetry regarding matter and antimatter in
the universe and it could also provide some insight on the problem
of dark matter. We believe that this issue merits further analysis.

\section{Conclusions and perspectives\label{Sec-Conclusions}}

In this work we discussed the magnetic effects of gravity according
to the special theory of relativity. A fully covariant theory of gravitation
with respect to the Lorentz transformations was constructed from first
principles: all results were derived from the special theory of relativity
only, making no use of any analogy with the electromagnetism theory
or additional assumptions. This covariant theory of gravitation, nevertheless,
share many of the properties of the electromagnetism theory, which
is due to requirements imposed by the special theory of relativity
for any relativistic force field. The present theory should be regarded
as an approximation of the general theory of relativity that holds
for bodies moving at large enough distances of the gravitational sources,
but not necessarily with small velocities. The Newtonian limit corresponds
to the case where the velocities are also very small when compared
with the speed of light.

The main conclusions derived in this work are the following: 
\begin{enumerate}
\item We showed through a simple thought experiment that any massive body
in motion should create a gravitational magnetic-like field. 
\item Imposing covariance of the gravitational force with respect to the
Lorentz transformations, we derived the exact expressions for these
gravitomagnetic fields. We showed as well that the total gravitational
force can always be written (for any inertial frame) in the same form
as the Lorentz force of the electromagnetism theory;
\item A comparison between the expression for the gravitational Lorentz
force derived here and those obtained from other approximations of
the general relativity was presented. We showed that linear approximations
of general relativity usually leads to a modified Lorentz force (e.g.,
introducing a factor of $4$ in the magnetic gravitational force)
and that this is due to the account of high order terms used in those
approximations. 
\item The differences between the concepts of mass (current) density and
energy (momentum) density were clarified. In particular, we showed
that the gravitational mass and current densities transform under
the Lorentz transformations as the components of a spacetime vector,
while the energy and momentum densities transform as two-rank tensor
components. 
\item This implies that the gravitational mass of a body must be an invariant,
in the same footing as the electric charge.
\item The differential equations satisfied by the gravitomagnetic fields
were also derived through covariance requirements only. They are similar
to the Maxwell equations of electromagnetism, except for the presence
of negative signs, which are due to the attractive character of the
gravitational interaction.
\item From these gravitational Maxwell equations, the existence of gravitational
waves was demonstrated. Our approach leads to a unique value for the
speed of gravitational waves, namely, the velocity of the light in
vacuum, $c$. We highlight that this result is in agreement with the
most recent experiments concerning the detection of gravitational
waves, which provides a strong argument in favor of the validity of
our theory in the approximation considered.
\item We introduced the gravitomagnetic potentials and we argued that a
physical meaning can be attached to them (as well as to their electromagnetic
analogues). Whenever a scalar potential is associated with an interaction
energy of the bodies, the corresponding vector potential should be
associated to a corresponding interaction momentum. The gauge freedom
of the potentials does not prohibit their reality as physical quantities,
since the ground state of the interaction energy and momentum can
also be chosen at will. 
\item Then, the energy and momentum stored on the gravitational fields were
discussed and a four-dimensional formulation of the theory was also
presented.
\item Finally, some differences between gravity and electricity were discussed.
The most important one comes from the always attractive behavior of
the gravitational interaction for usual massive bodies. We had speculated,
however, that this could not be so if a gravitational interaction
between particles and antiparticles is considered. 
\end{enumerate}
The present theory opens the door to the formulation of a quantum
theory of gravity that is expected to hold at large distances of the
gravitational sources, or in any situation where special relativity
can be used with safety. Such a theory could be relevant in discussing
some gravitational effects where quantum mechanics plays a significant
role, for instance, in the Hawking radiation, Landau levels, gravitational
entanglement, the gravitational analogues of the Aranov-Bohm, Casimir
and Unruh effects, among others. We expect to not find any mathematical
difficulty in formulating such a ``quantum gravitodynamics'' in a
flat spacetime background, since in this case the gravitomagnetism
is quite similar to the electromagnetism (being even gauge-invariant
also) and there is already a well-established formulation of quantum
electrodynamics (in fact, some work on this direction was presented
recently in \cite{BeheraBarik2017}). The conceptual issues, however,
might be challenging. For instance, it is usually believed that gravity
is mediated by the exchange of spin-2 particles \textendash{} the
so-called \emph{gravitons}. This is mainly because the gravitational
potentials in Einstein theory are described by a two-rank tensor \cite{FeynmanMorinigoWagner2003,Zee2010}.
However, we have seen here that the gravitational potentials can be
described as well by a spacetime vector in our approximation, in the
same way as the electromagnetic potentials. This means, therefore,
that we can also associate a spin-1 exchanging particle with the gravitational
interaction in the approximation considered. The connection between
this exchanging particle and the usual graviton, however, still need
to be clarified further. A related issue concerns with the conclusion
that spin-1 quantum field theory leads to a repulsive force between
likewise particles. This is usually taken as an argument to disprove
any attempt to construct a spin-1 theory of quantum gravity \cite{FeynmanMorinigoWagner2003,Zee2010}.
This argument, however, does not take into account that the energy
of the gravitational fields is negative, rather than positive as in
the electromagnetic case. This fact, however, makes likewise particles
to attract each other as they exchange spin-1 particles \textendash{}
i.e., the apparent counter-intuitive negativeness of the energy stored
in the gravitomagnetic fields is, notwithstanding, the ultimate reason
for why gravity is attractive, even on a quantum level. A detailed
exposition of these topics will be the subject of a future work.

\section*{Acknowledgments}

We thank Professor A. Lima-Santos for his advice given during the
development of these ideas and also Professor P. F. Farinas for the
reading of the manuscript and comments. A special thanks goes to Professor
A. De Rújula for providing us with his beautiful cartoon.

The work of RSV was supported by the São Paulo Research Foundation
(FAPESP), grants \#2012/02144-7 and \#2011/18729-1 and the Coordination
for the Improvement of Higher Education Personnel (CAPES). The work
of HBB was supported by the National Counsel of Technological and
Scientific Development (CNPQ), grants \#140911/2014-6 and \#870194/1997-6.

\appendix

\section{Conventions and notations}

In this paper, three-dimensional vectors are written in bold-italics
as in $\boldsymbol{A}$ or in the tensor notation as in $A^{a}$,
in which case Latin indices run from $1$ to $3$. Tensor quantities
defined on Minkowski spacetime are written in the usual tensor notation
as $A^{\mu}$, where Greek indices run from $0$ to $3$. We named
four-dimensional quantities by their spatial component name only,
preceded by the word ``spacetime'' (for example, we refer to the
\emph{spacetime position} $x^{\mu}$, the \emph{spacetime} \emph{momentum}
$p^{\mu}$, the spacetime \emph{momentum-flux} $T^{\mu\nu}$ and so
on).

We consider here the metric $\eta^{\mu\nu}=\text{diag}\left(1,-1,-1,-1\right)$
and the spacetime coordinates $x^{0}=ct$, $x^{1}=x$, $x^{2}=y$,
and $x^{3}=z$. Moreover, the Lorentz factor, 
\begin{equation}
\gamma(\boldsymbol{u})=\frac{1}{\sqrt{1-\boldsymbol{u}\cdot\boldsymbol{u}/c^{2}}}
\end{equation}
is usually denoted as $\gamma_{u}$, although $\gamma(v)$ is written
as $\gamma_{v}$ or just as $\gamma$ whenever it is clear from the
context. We write the usual partial derivatives as $\partial_{t}\equiv\partial/\partial t$,
$\partial_{x}\equiv\partial/\partial x$, $\partial_{y}\equiv\partial/\partial y$
and $\partial_{z}\equiv\partial/\partial z$, while the spacetime
derivatives are written as $\partial_{\mu}\equiv\partial/\partial x^{\mu}$
and $\partial^{\mu}\equiv\partial/\partial x_{\mu}$. Finally, the
\emph{dotted nabla} \emph{symbol}, 
\begin{equation}
\dot{\boldsymbol{\nabla}}=\hat{\boldsymbol{x}}\frac{\partial}{\partial\dot{x}}+\hat{\boldsymbol{y}}\frac{\partial}{\partial\dot{y}}+\hat{\boldsymbol{z}}\frac{\partial}{\partial\dot{z}},
\end{equation}
means the vector differential operator which takes derivatives w.r.t.
the velocity components.

Here we consider mainly three frames of reference: the frame $S'$
is a reference frame where a given distribution of mass (namely, that
one which generates the gravitational field), is always at rest. On
the other hand, the reference frame $S$ is regarded as a reference
frame where that distribution of mass moves with a constant velocity
$\boldsymbol{v}=v\hat{\boldsymbol{x}}$ (in other words, $S'$ moves
w.r.t. $S$ with the velocity $\boldsymbol{v}=v\hat{\boldsymbol{x}}$).
To avoid misunderstanding, in some situations we consider a third
reference frame $S''$, which is, in all aspects, the same as the
reference frame $S'$ except that no fixed distribution of mass is
attached to it. In fact, we assume that the velocity of that distribution
of mass is $\boldsymbol{u}$ w.r.t. $S$ and $\boldsymbol{u}''$ w.r.t.
$S''$. As usual, we assume further that all axes of these reference
frames are equally oriented and superposed at $t=t'=t''=0$.

\section{Vector identities}

The following vector identities are used in the paper: 
\begin{align}
\boldsymbol{\nabla}\cdot\left(\boldsymbol{\nabla}\times\boldsymbol{a}\right) & =0,\label{V1}\\
\boldsymbol{\nabla}\times\left(\boldsymbol{\nabla}f\right) & =0,\label{V2}\\
\partial_{t}\left(\boldsymbol{\nabla}\times\boldsymbol{a}\right) & =\boldsymbol{\nabla}\times\left(\partial_{t}\boldsymbol{a}\right),\label{V3}\\
\boldsymbol{\nabla}\times\left(\boldsymbol{\nabla}\times\boldsymbol{a}\right) & =\boldsymbol{\nabla}\left(\boldsymbol{\nabla}\cdot\boldsymbol{a}\right)-\boldsymbol{\nabla}^{2}\boldsymbol{a},\label{V4}\\
\boldsymbol{\nabla}\cdot\left(\boldsymbol{a}\times\boldsymbol{b}\right) & =\boldsymbol{b}\cdot\left(\boldsymbol{\nabla}\times\boldsymbol{a}\right)-\boldsymbol{a}\cdot\left(\boldsymbol{\nabla}\times\boldsymbol{b}\right),\label{V5}\\
\partial_{t}\left(\boldsymbol{a}\times\boldsymbol{b}\right) & =\left(\partial_{t}\boldsymbol{a}\right)\times\boldsymbol{b}+\boldsymbol{a}\times\left(\partial_{t}\boldsymbol{b}\right),\label{V6}\\
\boldsymbol{a}\times\left(\boldsymbol{\nabla}\times\boldsymbol{a}\right) & =\tfrac{1}{2}\boldsymbol{\nabla}\left(\boldsymbol{a}\cdot\boldsymbol{a}\right)-\left(\boldsymbol{a}\cdot\boldsymbol{\nabla}\right)\boldsymbol{a},\label{V7}\\
\boldsymbol{\nabla}\left(\boldsymbol{a}\cdot\boldsymbol{b}\right) & =\left(\boldsymbol{a}\cdot\boldsymbol{\nabla}\right)\boldsymbol{b}+\left(\boldsymbol{b}\cdot\boldsymbol{\nabla}\right)\boldsymbol{a}+\boldsymbol{a}\times\left(\boldsymbol{\nabla}\times\boldsymbol{b}\right)+\boldsymbol{b}\times\left(\boldsymbol{\nabla}\times\boldsymbol{a}\right),\label{V8}
\end{align}
where $f$ is an at least $C^{2}$ class scalar function and $\boldsymbol{a}$,
$\boldsymbol{b}$ and $\boldsymbol{c}$ are at least $C^{2}$ vector
functions. We also make use of the following tensor identity: 
\begin{equation}
\boldsymbol{\nabla}\cdot\left(\boldsymbol{a}\otimes\boldsymbol{a}\right)=\left(\boldsymbol{\nabla}\cdot\boldsymbol{a}\right)\boldsymbol{a}+\left(\boldsymbol{a}\cdot\boldsymbol{\nabla}\right)\boldsymbol{a}.\label{VT}
\end{equation}

\bibliographystyle{spiebib}
\bibliography{GM}

\end{document}